\documentclass[aps,reprint,showpacs,twocolumn,superscriptaddress,floatfix]{revtex4-2}

\usepackage{bm}
\usepackage{graphicx}
\usepackage{physics}

\def\be{\begin{equation}}
\def\ee{\end{equation}}
\def\beq{\begin{eqnarray}}
\def\eeq{\end{eqnarray}}

\begin{document}
\author{Fabio Cavaliere}
    \affiliation{Dipartimento di Fisica, Universit\`a di Genova, Via Dodecaneso 33, 16146 Genova, Italy} 
    \affiliation{CNR-SPIN,  Via  Dodecaneso  33,  16146  Genova, Italy}
\email{fabio.cavaliere@fisica.unige.it}
\author{Matteo Carrega}
\affiliation{CNR-SPIN,  Via  Dodecaneso  33,  16146  Genova, Italy}
\author{Giulio De Filippis}
    \affiliation{SPIN-CNR and Dip. di Fisica ``E.~Pancini'' - Universit\`a di Napoli ``Federico II'' - I-80126 Napoli, Italy.}
     \affiliation{INFN, Sezione di Napoli, Complesso Universitario di Monte S. Angelo,
I-80126 Napoli, Italy}
\author{Vittorio Cataudella}
    \affiliation{SPIN-CNR and Dip. di Fisica ``E.~Pancini'' - Universit\`a di Napoli ``Federico II'' - I-80126 Napoli, Italy.}
\author{Giuliano Benenti}
    \affiliation{Center for Nonlinear and Complex Systems, Dipartimento di Scienza e Alta Tecnologia, Universit\`a degli Studi dell'Insubria, via Valleggio 11, 22100 Como, Italy} 
    \affiliation{Istituto Nazionale di Fisica Nucleare, Sezione di Milano, via Celoria 16, 20133 Milano, Italy}
    \affiliation{NEST, Istituto Nanoscienze-CNR, I-56126 Pisa, Italy}
    \author{Maura Sassetti}
    \affiliation{Dipartimento di Fisica, Universit\`a di Genova, Via Dodecaneso 33, 16146 Genova, Italy} 
    \affiliation{CNR-SPIN,  Via  Dodecaneso  33,  16146  Genova, Italy}

\title{Dynamical heat engines with non--Markovian reservoirs}
\date{\today}
\begin{abstract}
We discuss whether, and under which conditions, it is possible to realize a heat engine simply by dynamically modulating the couplings between the quantum working medium and thermal reservoirs.
For that purpose, we consider the paradigmatic model of a quantum harmonic oscillator, exposed to a minimal modulation, that is, a monochromatic driving of the coupling to only one of the thermal baths. 
We demonstrate, at any order in the system/bath coupling strength, that in this setup non--Markovianity of the bath is a necessary condition to obtain a heat engine.
In addition, we identify suitable structured environments for the engine to approach the ideal  Carnot efficiency.
Our results open up new possibilities for the use of non--Markovian open quantum systems for the construction and optimization of quantum thermal machines.
\end{abstract}
\maketitle
\section{Introduction}
\label{sec:intro}
The trend toward miniaturization is pushing heat engines up to the level where the working medium is a small system, which requires quantum mechanics for an accurate description~\cite{pekola15, vinjanampathy2015, benenti17, curzon, campaiolibook, killoran, ciliberto, bouton, thierschmann15, martinez16, blasi21, xu22}. This opened several fundamental and applicative issues in the growing field of quantum thermodynamics~\cite{esposito09, campisi11, kosloff13, goold2016, carrega19, miller19, khandelwal, vischi19, bhandari21, pancotti20, jurgen, liu22, esposito19, brandner166, son22} in the last years. As for any other quantum machine, the  interaction of the quantum system with the external world requires special care~\cite{vinjanampathy2015, curzon, xu22, pancotti20, jurgen, breuerbook, weiss, aurel18, strasberg16, restrepo18}. On one hand, it would be desirable to maintain the system isolated from the environment, to preserve any quantum advantage~\cite{campaiolibook, qcbook, wang, elouard, rossini, gyhm, watanabe0, watanabe, hammam} provided
by coherent dynamics. On the other hand, a thermal engine delivering finite power requires that work is extracted and heat is exchanged with reservoirs at finite rates. A rigorous treatment of energy exchanges and heat flows is thus required to properly model quantum thermal machines working out of equilibrium.
For instance, the coupling, possibly strong~\cite{liu22, talkner, moskalets, paternostro}, between quantum working medium and baths, can quite naturally induce non--Markovian effects~\cite{devega, landi_prxquantum, hewgill, carrega_prxquantum, ivander22, shirai, breuerrmp, nori12, newpt}, which are often either  not captured or discarded by common approximation schemes~\cite{vinjanampathy2015, benenti17, curzon, kosloff13, bhandari21, hewgill, leitch, hofer}.
The question then arises, whether non--Markovianity may constitute a useful thermodynamic resource. 

In this work, we address this question for a minimal disturbance of the quantum system, that is, a monochromatic modulation of the coupling to one thermal bath. The same achievement of a heat engine in such a setup, without directly driving the system, is a non trivial result. Indeed, modulation of the couplings is intuitively associated with dissipation, akin to friction induced by moving parts in 
strokes of a macroscopic heat engine. Non--Markovianity, associated to the spectral properties of the bath, is here investigated in the paradigmatic model of a quantum harmonic oscillator (QHO)~\cite{nori12, haake, hu92}, coupled to two bosonic thermal (hot and cold) baths. 
Such approach is quite versatile, since it is possible to study the QHO dynamics and thermodynamics, without resorting to any approximations, both in the quantum and in the classical regime, and for arbitrary spectral features of the environment.  

We show that, as counterintuitive as it might seem, a dynamical heat engine can be obtained in the above configuration. To achieve such a result we demonstrate that non--Markovianity inherited from the reservoir that feels the driven contact is a necessary but not sufficient condition. Furthermore, we show that by taking advantage of a suitable structured environment, the engine can even approach the Carnot efficiency.\\

The paper is organized as follows: in Sec.~\ref{sec:generalfw} we introduce the model and outline the methods employed to evaluate the average power and heat currents. The connections between non--Markovianity and a working heat engine are discussed in Sec.~\ref{sec:dynheat}. Subsequently, exploiting the prototypical non--Markovian bath with a Lorentzian spectral density we discuss in Sec.~\ref{sec:NMengine} the performances of the ensuing heat engine, both in the weak and in the strong coupling regime. An interpretation of the weak coupling limit in terms of quantum Otto cycles is also provided. Conclusions are finally drawn in Sec.~\ref{sec:conclusions}.

\section{General framework}
\label{sec:generalfw}
\subsection{Model}
\label{subsec:model}
The working medium of the thermal machine is a QHO whose Hamiltonian  reads ($\hbar=k_{{\rm B}}=1$) 
\begin{equation}
\label{eq:HQHO}
H_{{\rm QHO}}= \frac{p^2}{2m} + \frac{1}{2}m\omega^2_{0} x^2\,, 
\end{equation}
where $m$ and $\omega_0$ are its mass and characteristic
frequency, respectively. The QHO is linearly coupled to two reservoirs, with the total Hamiltonian
\begin{equation}
    \label{eq:Htott}
H^{(t)}  = H_{{\rm QHO}} + \sum_{\nu=1}^2 \left(H_\nu + H^{(t)}_{{\rm int},\nu}\right)\,.
\end{equation}
Each bath ($\nu=1,2$) is modeled as an ensemble of harmonic oscillators in the usual Caldeira-Leggett~\cite{breuerbook, weiss, aurel18, CL83, cangemi} framework with Hamiltonians
\begin{equation}
    \label{eq:Hnu}
H_{{\nu}}=  \sum_{k=1}^{\infty} \left(\frac{P^2_{k,\nu}}{2 m_{k,\nu}} + \frac{m_{k,\nu} \omega^2_{k,\nu }X^2_{k,\nu}}{2}\right)\,.
\end{equation}
We assume that the system/baths couplings can be varied in time~\cite{jurgen, carrega_prxquantum, flindt21}, described by the interaction contribution 
\begin{equation}
    \label{eq:Hint}
\!\!\!\!H^{(t)}_{{\rm int,\nu}}\!\!=\!\!\sum_{k=1}^{\infty}\left\{-x g_\nu(t)c_{k,\nu}X_{k,\nu}\!+\!x^2g_{\nu}^2(t)\frac{c^2_{k,\nu}}{2m_{k,\nu} \omega^2_{k,\nu}}\right\}\,.
\end{equation}
The interaction strengths are described by the parameter $c_{k,\nu}$, and the time-dependence of the couplings is in the dimensionless periodic functions $g_\nu(t)$ satisfying $g_{\nu}(t)=g_{\nu}(t+\mathcal{T})$ with Fourier decomposition
\begin{equation}
    \label{app_seriesegSPr}
    g_{\nu}(t)=\sum_{n=-\infty}^{\infty}g_{n,\nu}\ e^{-in\Omega t}\,;\,\Omega=\frac{2\pi}{\mathcal{T}}\, .
\end{equation}
\begin{figure}
\includegraphics[width=8cm,keepaspectratio]{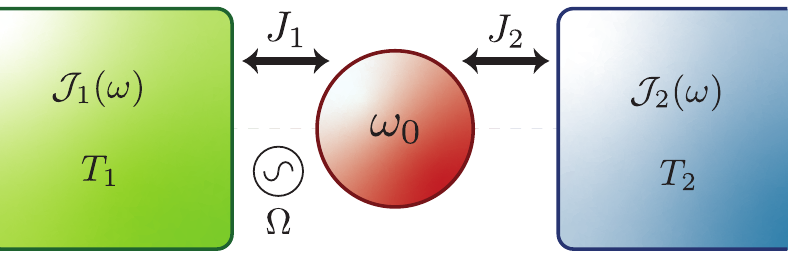}
\caption{Sketch of the setup: $J_\nu$ represents the energy current flowing between the QHO and the $\nu$--th contact. The two reservoirs are in equilibrium at a temperature $T_\nu$ and described by a spectral density ${\cal J}_\nu(\omega)$ with $\nu=1,2$. The coupling of the bath $\nu=1$ is modulated by a monochromatic driving with frequency $\Omega$, while the coupling to the bath $\nu=2$ is static.}\label{fig:fig1}	
\end{figure}
In this paper we consider the minimum modulation needed for the couplings in the search for a heat engine, that is a monochromatic drive at frequency $\Omega$ for
the first contact while the second is kept constant
\begin{equation}
\label{eq:drivingscheme}
g_1(t)=\cos(\Omega t)\quad;\quad g_2(t)=1\,,
\end{equation}

\noindent see the sketch in Fig. \ref{fig:fig1}. To model the bath properties we introduce their spectral densities~\cite{weiss}
\be
\label{eq:spectral}
{\cal J}_\nu(\omega)=\frac{\pi}{2}\sum_{k=1}^\infty \frac{c_{k,\nu}^2}{m_{k,\nu}\omega_{k,\nu}}\delta(\omega-\omega_{k,\nu})~,
\ee
whose precise forms will be specified later. 

At the initial time $t_0{\to -\infty}$ the baths are assumed in their thermal equilibrium at temperatures $T_\nu$, with the total  density matrix, written in a factorized form $\rho(t_0) = \rho_{\rm QHO}(t_0)\otimes \rho_{1}(t_0)\otimes 
\rho_{2}(t_0)$, where $\rho_{\rm QHO}(t_0)$ is the initial system density matrix. 

The out of equilibrium dynamic of the QHO obeys the generalized quantum Langevin equation~\cite{carrega_prxquantum, paz1, paz2}
\begin{widetext}
\begin{equation}\label{eq:diffqoper}
\ddot{x}(t) + \omega^2_0 x(t)+\int_{t_0}^{+\infty}\!\!\mathrm{d}s\sum_{\nu=1}^2 g_\nu(t)\gamma_\nu(t-s) \Big[\dot{g}_\nu(s)x(s)+
\dot{x}(s)g_\nu(s)\Big] =\frac{1}{m}\sum_{\nu=1}^2 g_\nu(t)\xi_\nu(t)~,
\end{equation}
\end{widetext}
where overdots denote time derivatives and the damping kernels $\gamma_{\nu}(t)$ are linked to the spectral function by
\begin{equation}\label{eq:gamma}
\gamma_\nu(t)=\frac{2}{\pi m}\theta(t)\int_0^{\infty}\mathrm{d}\omega \frac{{\cal J}_\nu(\omega)}{\omega}\cos(\omega t),
\end{equation} 
with $\theta(t)$ the Heaviside step function. The fluctuating force operators $\xi_\nu(t)$ to the right hand side of Eq.~\eqref{eq:diffqoper} explicitly depend on the initial values of the bath operators $X_{k,\nu}(t_0)$ and $P_{k,\nu}(t_0)$. Their expression is
\begin{eqnarray}
\xi_\nu(t)&=&\sum_{k=1}^{\infty}c_{k,\nu}\Big[X_{k,\nu}(t_0)\cos\omega_{k,\nu} (t-t_0)\nonumber\\
&+&\frac{P_{k,\nu}(t_0)}{m_{k,\nu}\omega_{k,\nu}}\sin\omega_{k,\nu} (t-t_0)\Big]\,.\label{app_randomFSP}
\end{eqnarray}

We recall that these operators have zero quantum average $\langle \xi_\nu(t)\rangle\equiv {\rm Tr}[\xi_\nu (t) \rho(t_0)]=0$ and their time correlators are given by $\langle \xi_\nu (t)\xi_{\nu'}(t')\rangle =\delta_{\nu,\nu'}{\cal L}_\nu (t-t')$, with
\be
\label{correlator}
\!\!\!\!\!\!{\cal L}_\nu (t)\!\!=\!\!\!\int_0^{\infty}\!\!\frac{\mathrm{d}\omega}{\pi} 
{\cal J}_\nu(\omega)\big[\!\coth(\frac{\omega}{2T_\nu})\!\cos(\omega t)-\!i \sin(\omega t)\big].
\ee

To solve the Langevin Equation~\eqref{eq:diffqoper} one needs the retarded Green's function ${G}(t,t')$, which obeys the integro--differential equation

\begin{eqnarray}
&&\ddot{G}(t,t') + \omega_0^2 G(t,t')+\int_{t_0}^{+\infty}\!\!\mathrm{d}s\sum_{\nu=1}^{2}g_\nu(t)\gamma_\nu(t-s)\nonumber\\
&&\times[\dot{g}_\nu(s)G(s,t')+
g_\nu(s)\dot{G}(s,t')] =\delta(t-t'),\label{app_greenSP}\,.
\end{eqnarray}

At long times $t$, when the memory of the initial state for the QHO is lost, the time evolution of the position operator $x(t)$ is directly expressed  as a time integral of the retarded Green function as:
\be
\label{app_solutionSP}
x(t)=\frac{1}{m}\lim_{t_0\to-\infty}\int_{t_0}^{+\infty}\mathrm{d}t' G(t,t')\sum_{\nu=1}^{2}g_\nu(t')\xi_\nu(t')~\,.
\ee

As we will see shortly, the key relation Eq.~(\ref{app_solutionSP}) will allow us to evaluate all quantum correlation averages, associated to thermodynamic observables.  Notice that at long times $G(t,t')$ acquires the following peculiar form:  
\be
\label{app_transform}
G(t,t')=\sum_{\mu=-\infty}^{+\infty}\int_{-\infty}^{+\infty}\frac{\mathrm{d}\omega}{2\pi}e^{-i\omega(t-t')}\tilde{ G}_\mu(\omega) e^{-i\mu\Omega t},\ee
where $\tilde{ G}_{\mu}(\omega)$  are the so-called  Floquet coefficients obeying the following set of algebraic equations~\cite{carrega_prxquantum}:
\begin{eqnarray}
\tilde{ G}_\mu(\omega)&=&\chi(\omega)\delta_{\mu,0}-\chi(\omega+\mu\Omega)\nonumber\\
&&{\times}\sum_{n=\pm 2}\tilde{k}_n(\omega+(\mu-n)\Omega) \tilde{ G}_{\mu-n}(\omega).\label{app_sfloquet1}
\end{eqnarray}
Here we have introduced
\begin{equation}
\label{eq:chimain}
{\chi}(\omega)=-\frac{1}{\omega^2-\omega_0^2-\tilde{k}_0(\omega)}\,,
\end{equation}
and
\begin{equation}
\label{app_chikappaSP}
\!\!\!\!\tilde{k}_n(\omega)\!\!=\!\!-i\sum_{\nu=1}^2 \sum_{\mu=-\infty}^{+\infty}g_{\mu,\nu}g_{n-\mu,\nu}(\omega+\mu\Omega)\tilde\gamma_\nu(\omega+\mu\Omega),
\end{equation}
with $\tilde{\gamma}_{\nu}(\omega)$ the Fourier transform of $\gamma_{\nu}(t)$.\\

\subsection{Average thermodynamic quantities}
\label{subsec:thermodyn}
We study thermodynamic quantities in the long time limit, when a periodic steady state has been reached. Notice that the study of instantaneous quantities and their time dependence should be considered with care, especially at finite coupling strength ~\cite{jurgen,moskalets,liu22,esposito19}: this however is a problem outside the scope of our paper. We are in particular interested in quantities {\em averaged over one period of the cycle}, that are well defined both in the weak and in the strong coupling regime~\cite{jurgen,esposito19,liu22, brandner166}. 

We start by considering the operatorial time evolution of the total Hamiltonian~(\ref{eq:Htott}) in the Heisenberg representation:
\be
\label{app_powerSP}
\!\!\!\!\frac{d}{dt}\left[\sum_{\nu=1}^{2} H_\nu (t)+H_{{\rm QHO}}(t) + H_{{\rm int}}^{(t)}(t)\right]= \frac{\partial}{\partial t}H_{{\rm int}}^{(t)}(t)~,
\ee
where $H_{\mathrm{int}}^{(t)}(t)=\sum_{\nu=1}^{2}H_{\mathrm{int},\nu}^{(t)}(t)$ -- see Eq.~\eqref{eq:Hint}.

Taking the quantum ensemble average, and performing the average over one period of the cycle, we obtain
\be
P+\sum_{\nu=1}^2 J_\nu+\mathcal{A}=0,
\ee
where we have introduced
\begin{widetext}
\beq
&&P=\frac{1}{\cal T}\int_t^{t+{\cal T}}\!\!\!\!\! dt'\,\sum_{\nu=1}^2{\rm Tr}
\Big[\frac{\partial H_{{\rm int},\nu}^{(t')}(t')}{\partial t'}\rho(t_0)\Big]=\frac{1}{\cal T}\int_t^{t+{\cal T}}\!\!\!\!\!dt'\,\sum_{\nu=1}^2{\rm Tr}
\Big[\frac{\partial H_{{\rm int},\nu}^{(t')}}{\partial t'}\rho(t')\Big],
\label{app_PSP}\\
&& J_\nu=-\frac{1}{\cal T}\int_t^{t+{\cal T}}\!\!\!\!\! dt'\,{\rm Tr}
\Big[\frac{d}{dt'}H_{\nu}(t')\rho(t_0)\Big]=-\frac{1}{\cal T}\int_t^{t+{\cal T}}\!\!\!\!\! dt'\,{\rm Tr}
\Big[H_{\nu}\frac{d}{dt'}\rho(t')\Big],\label{app_JSP}\\
&&{\cal A}=\frac{1}{\cal T}\int_t^{t+{\cal T}}\!\!\!\!\!dt'\,{\rm Tr}\Big[
\frac{d}{dt'}\Big(H_{\rm QHO}(t')+H_{\rm int}^{(t')}(t')\Big)\rho(t_0)\Big]~.\label{app_ASP}
\eeq
\end{widetext}
Here, $P$ is the total power associated to the time evolution of the system/bath couplings, $J_{\nu}$ is the current energy flow from the  $\nu$-th reservoir, and ${\cal A}$ represents the remaining contributions stemming from the QHO and the interaction term. It has been argued that, in general, the term $\mathcal{A}$ can be nonzero~\cite{liu22}. However, in our case we can show -- see Appendix~\ref{appendix1} -- that ${\cal A}=0$. This important result implies that in the long time limit and after the cycling average the total power due to the coupling drives is totally balanced by the reservoir heat currents and fulfills  the relation 
\be
 \label{app_firstlaw1}
P +\sum_{\nu=1}^2J_\nu=0\,,
\ee
which can be interpreted as a manifestation of the first law of thermodynamics. To actually evaluate the average power $P$ and heat currents $J_{\nu}$ at periodic steady state, we resort to a non--equilibrium Green function formalism~\cite{carrega_prxquantum, paz1, paz2, arrachea12a, grifoni95}. Deferring all details to Appendix~\ref{appendix2}, here we quote the final results for the power and the heat current via bath $\nu=2$ (recall that $J_1=-P-J_2$)
\begin{widetext}
\beq
\label{Paverage}
&&P=\frac{\Omega}{4\pi m} \sum_{\mu=-\infty}^{+\infty} \int_{-\infty}^{+\infty}\!\!\mathrm{d}\omega\Big\{\frac{{\cal J}_{2}(\omega)}{2m}\coth(\frac{\omega}{2T_{2}})\!\Big|\tilde{ G}_{\mu}(\omega)\Big|^2
\Big[{\cal J}_{1}(\omega+(\mu+1)\Omega)\!-\!{\cal J}_{1}(\omega+(\mu-1)\Omega)\Big]\nonumber\\
&&+{\cal J}_1(\omega)\coth(\frac{\omega}{2T_1}) \Big\{\frac{{\cal J}_{1}(\omega-\mu\Omega)}{4m}|\tilde{ G}_{\mu}(-\omega+\Omega)+\tilde{ G}_{\mu+2}(-\omega-\Omega)|^2-\Im\left[\tilde{ G}_{0}(\omega-\Omega)\right] \delta_{\mu,0}\Big\}\Big\},
\eeq
\beq
\label{app_sjaaverage}
&&J_2=\frac{1}{2\pi m} \sum_{\mu=-\infty}^{+\infty} \int_{-\infty}^{+\infty}\!\!\mathrm{d}\omega\Big\{ {\cal J}_2(\omega)\coth(\frac{\omega}{2T_2})\Big[\omega \Im\left[\tilde{ G}_{0}(\omega)\right] \delta_{\mu,0}-\frac{{\cal J}_{2}(\omega-\mu\Omega)}{m}
(\omega -\mu \Omega) \Big|\tilde{ G}_{\mu}(\omega)\Big|^2\Big] +\nonumber \\
&&-\frac{{\cal J}_{1}(\omega)}{4m}\coth(\frac{\omega}{2T_{1}}) (\omega-\mu \Omega) {\cal J}_{2}(\omega-\mu\Omega)
\Big|\tilde{ G}_{\mu-1}(-\omega+\Omega) +\tilde{ G}_{\mu+1}(-\omega-\Omega)\Big|^2\Big\}\,,
\eeq
\end{widetext}
where the influence kernels have the form 
\be
\label{kappan}
\!\!\tilde{k}_0(\omega)=-i\omega\tilde{\gamma}_2(\omega)+\!\!\!\sum_{n=\pm2}\!\!\!\tilde{k}_{n}(\omega)\ ;\ 
\tilde{k}_{\pm2}(\omega)=-\frac{i}{4} \omega_{\pm}\tilde{\gamma}_1 (\omega_{\pm}),
\nonumber
\ee
induced by the specific driving scheme considered in this work -- see Eq.~(\ref{eq:drivingscheme}). We close noting that the Floquet coefficients possess only even $\mu$ components and have the following symmetry properties: $\tilde{G}_\mu^*(\omega)=\tilde{G}_{-\mu}(-\omega)$ and $\tilde{ G}_{\mu}(\omega-\frac{\mu}{2}\Omega)=\tilde{ G}_{-\mu}(\omega+\frac{\mu}{2}\Omega)$ -- see Appendix.~\ref{appendix3} for the latter.\\

\section{Dynamical heat engine versus Markovianity}
\label{sec:dynheat}
In general, a structured environment can induce memory effects and non--Markovian dynamics. To assess non--Markovianity, several estimators have been introduced recently~\cite{breuerrmp, rivas10, groeblacher, illuminatiprl}. As suggested in Refs.~\cite{groeblacher, illuminatipra}, a proper criterion to quantify non--Markovianity in the asymptotic regime is the violation of divisibility property of the dynamical map. Moreover, a direct link between the non--divisibility notion of non--Markovianity and the form of the bath spectral density has been given. In particular, it is possible to show~\cite{groeblacher, illuminatipra} that \emph{ only} a strictly Ohmic~\cite{noteOhmic} spectral density ${\cal J} (\omega) = m\gamma \omega$ in the classic (high temperature) regime leads to a separable map, hence to a Markovian dynamics -- see Appendix~\ref{appendix4} for details.\\ 

We now inspect general  and necessary conditions to reach a  heat engine regime, that is $P<0$. By studying Eq.~(\ref{Paverage}), as discussed in Appendix~\ref{appendix2}, one can realize that ${\cal J}_1(\omega)$, i.e. the bath spectral density linked to the driven contact, dictates the sign of the different contributions of the average power.\\ 

We begin considering a strictly Ohmic spectral function
\begin{equation}
\label{eq:strictohm}    
{\cal J}_1(\omega)=m\gamma_1\omega
\end{equation}
in the high temperature regime. 
We now show that {\em this choice, which represents a Markovian dynamics at high temperature, cannot support a heat engine at any order in the system/bath interaction} in the high temperature regime. As shown in Appendix~\ref{appendix2}, the average power $P$ can be decomposed as
\begin{equation}
\label{eq:decomp}
P=P^{(a)}+P^{(b)},\ \ \mathrm{where}\ P^{(b)}=P^{(b,1)}+P^{(b,2)}\,.
\end{equation}
Inserting Eq.~(\ref{eq:strictohm}) into Eqs.~\eqref{app_p1afinal},~\eqref{app_P1b1final} and~\eqref{app_P1b2final} an explicit expression for the above three quantities is obtained.
We thus arrive at
\be
\label{app_P1aohm}
\!\!\!\!\!\!\!\!P^{(a)}=-\frac{\Omega\gamma_1}{4\pi}\int_{-\infty}^{+\infty}\!\!\mathrm{d}\omega\omega\coth(\frac{\omega}{2T_1})\Im\left[\tilde{ G}_{0}(\omega-\Omega)\right],
\ee
\begin{eqnarray}
\!\!\!\!\!\!\!\!\!\!\!\!\!\!\!\!P^{(b,1)}&=&\frac{\Omega\gamma_1^2}{16\pi}\int_{-\infty}^{+\infty}\mathrm{d}\omega\omega\coth(\frac{\omega}{2T_{1}})\nonumber\\
\!\!\!\!\!\!\!\!\!\!\!\!\!\!\!\!\!\!\!\!\!\!\!\!&\!\!\!\!\!\!\!\!\!\!\!\!\!\!\!\!\!\!\!\!\times&\!\!\!\!\!\!\!\!\!\!\!\sum_{\mu=-\infty}^{+\infty}(\omega-\mu\Omega) |\tilde{ G}_{\mu}(-\omega+\Omega)+\tilde{ G}_{\mu+2}(-\omega-\Omega)|^2,\label{app_P1b1ohm}
\end{eqnarray}
and
\be
\label{app_P1b2ohm}
P^{(b,2)}\!\!=\frac{\Omega^2\gamma_1}{4\pi m}\!\!\!\int_{-\infty}^{+\infty}\!\!\!\!\!\!\!\mathrm{d}\omega{\cal J}_2(\omega)\coth(\frac{\omega}{2T_{2}})\!\!\!\!\sum_{\mu=-\infty}^{+\infty}\!\!\!|\tilde{ G}_{\mu}(\omega)|^2.
\ee
It is now easy to see that $P^{(b,2)}\ge 0$. It is also possible to show that $P^{(b,1)}\ge 0$. Indeed, by using Eq.~\eqref{app_property1} we first note that
\begin{eqnarray}
&&\int_{-\infty}^{+\infty}\mathrm{d}\omega\omega\coth(\frac{\omega}{2T_{1}})\nonumber\\
\!\!\!\!\!\!\!\!\!\!\!\!&\!\!\!\!\!\!\!\!\!\!\!\!\!\!\!\!\times\!\!\!\!\!\!\!\!&\!\!\!\!\!\!\!\!\sum_{\mu=-\infty}^{+\infty}\omega 
|\tilde{ G}_{\mu}(-\omega+\Omega)+\tilde{ G}_{\mu+2}(-\omega-\Omega)|^2=0\,.\label{app_P1b1ohm1}
\end{eqnarray}
The remaining part of $P^{(b,1)}$ can then be rewritten, after some algebra, as
\begin{eqnarray}
P^{(b,1)}&=&\frac{\Omega^2\gamma_1^2}{16\pi}\int_{-\infty}^{+\infty}\mathrm{d}\omega \omega\coth(\frac{\omega}{2T_1})\nonumber\\\
\!\!\!&\!\!\!\times&\!\!\!\sum_{\mu = -\infty}^{+\infty} |\tilde{G}_\mu (-\omega + \Omega) + \tilde{G}_{\mu + 2}(-\omega - \Omega) |^2\,,\label{app_p1b1w3}
\end{eqnarray}
which is also manifestly positive. This implies that $P^{(b)}\geq 0$, and that the possibility to have a heat engine only depends on the sign of $P^{(a)}$. Let us now consider the Markovian limit of high temperatures $T_1\to\infty$. Here $\coth(\frac{\omega}{2T_1})\!\to\!\frac{2T_1}{\omega}$, then using also the odd parity property of  $\Im[\tilde{ G}_{0}(\omega)]$ we find $P^{(a)}=0$.

This finally proves that {\em in the Markovian regime}{ $P>0$ to every order in the system/bath coupling strength}, demonstrating that in this case it is {\it not } possible to obtain a working heat engine ($P<0$). Thus, non--Markovianity is a necessary condition to achieve a working heat engine. 

It is worth to stress that the above argument holds true independently from the shape of ${\cal J}_2(\omega)$, related to the static bath.\\

It is now natural to wonder if this is also a sufficient condition. However, this is not the case, as we now argue by providing a counterexample. To this end, we consider the case in which the system/bath coupling strength of the modulated in time reservoir ($\nu=1$) is weak. In this perturbative regime, simpler closed expressions are obtained, from which one can also get useful physical intuitions. Up to linear order in ${\cal J}_1(\omega)$ the average heat power can be written as (see Appendix~\ref{appendix5} for details, in particular Eq.~(\ref{app_eq:Ppert}))
\be
\label{finalP}
P=-\Omega\int_{0}^{+\infty}\!\!\frac{\mathrm{d}\omega}{2\pi m}\Im\chi_0(\omega)f(\omega, \Omega).
\ee
Here, we introduced $\chi_0(\omega)=-[\omega^2-\omega_0^2 + i \omega\tilde{\gamma}_2(\omega)]^{-1}$ the bare susceptivity and
\begin{eqnarray}
&&f(\omega,\Omega)=  {\cal J}_1 (\omega_+) n_B \left(\frac{\omega_+}{T_1} \right) -
  {\cal J}_1 (\omega_-) n_B \left(\frac{\omega_-}{T_1} \right)\nonumber\\ 
&&+[{\cal J}_1 (\omega_-)-{\cal J}_1 (\omega_+)] n_B \left(
  \frac{\omega}{T_2} \right) ,\label{f_def}
\end{eqnarray}
where we recall that $\omega_{\pm}=\omega\pm\Omega$ and where $n_B(x)= (e^x -1)^{-1}$ is the Bose distribution function. Note that ${\cal J}_2(\omega)$ only enters into the expression of $\chi_0(\omega)$ through $\tilde{\gamma}_2(\omega)$~\cite{notechi}. Recalling that ${\cal J}_2(\omega)=m\omega \Re [\tilde{\gamma}_2(\omega)]$, one can realize that $\Im[\chi_0(\omega)]$ is positive for $\omega>0$. Therefore, the regions of a working heat engine are given by
\begin{equation}
f(\omega, \Omega)>0\,.\label{eq:thecondition}
\end{equation}
To look for a counterexample, for the modulated bath we consider a monotonically increasing spectral density of the form ${\cal J}_1(\omega)=m\gamma_1 \omega |\frac{\omega}{\bar{\omega}}|^{s-1}$, which describes a large class of spectral function: Ohmic behavior for $s=1$, sub-Ohmic for $0<s<1$, and super-Ohmic  for $s>1$~\cite{weiss, noteOhmic}. In this case, the last sum that appears in square brackets on the second line of Eq.~\eqref{f_def} is always negative, then the most favorable requirement for a working heat engine is in the $T_2\to 0$ limit, where the second
line vanishes. Then, taking advantage of the relation $\tilde{{\cal L}}_1(-\omega)=2{\cal J}_1(\omega) n_B(\omega/T_1)$, with $\tilde{\cal L}_1(\omega)$ Fourier transform of the fluctuating force correlator ${\cal L}_1(t)$, the condition for a working heat engine reads $\tilde{\cal L}_1(-\omega-\Omega)>\tilde{\cal L}_1(-\omega+\Omega)$. In passing, this confirm the necessary condition of non--Markovianity, since a Markovian bath has ${\cal L}_1(t)\propto \delta(t)$, and it never satisfies the above constraint~\cite{notejw}. More importantly, this relation can be used to obtain the counterexample we are looking for: Indeed, after some lengthy calculations reported in Appendix~\ref{appendix6}, it is possible to demonstrate that in the non--Markovian case of sub--Ohmic spectral density no heat engine can be achieved. Therefore, we conclude that \emph{non--Markovianity is a necessary but not sufficient condition to obtain a heat engine}.\\

\section{non--Markovian engine with a Lorentzian bath}
\label{sec:NMengine}
Having established the importance of non--Markovianity for a dynamical heat engine, we now characterize its performance. To this end, hereafter we choose a strictly Ohmic~\cite{noteOhmic} spectral density ${\cal J}_2(\omega)=m\gamma_2\omega$ for the $\nu=2$ static bath. For the modulated bath of interest, we focus on a paradigmatic example of structured non--Markovian environment, i.e. a Lorentzian spectral function~\cite{strasberg16, restrepo18, thorwart, paladino, nazir14} 
\be
{\cal J}_1(\omega)= \frac{d_1 m\gamma_1 \omega}{(\omega^2 - {\omega}_1^2)^2 + \gamma_1^2\omega^2}~,
\ee
with a peak centered at ${\omega}_1$, an amplitude governed by  $d_1$, and a width determined by $\gamma_1$ (parameter linked to the damping).
\begin{figure}
\includegraphics[width=8cm,keepaspectratio]{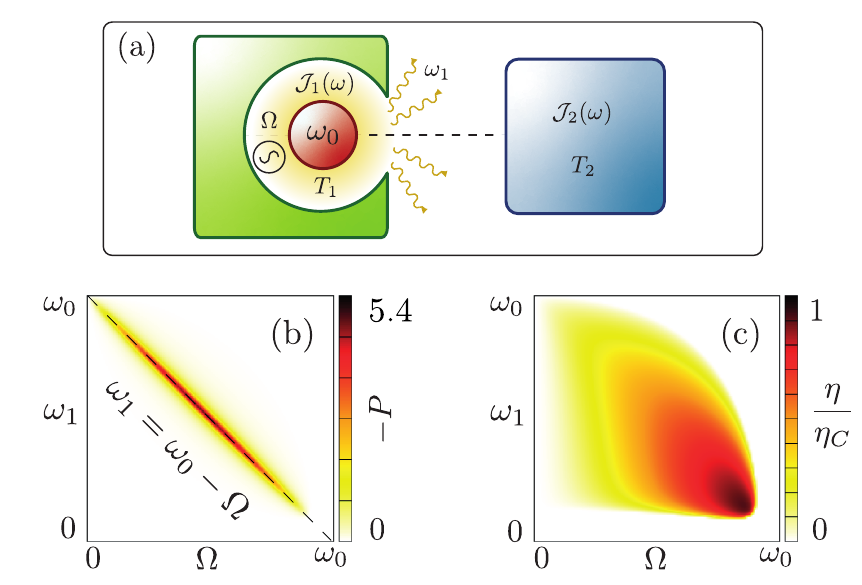}
\caption{Heat engine with a structured environment as pictorially depicted in panel (a). Panels (b) and (c) respectively show the engine average power (in unit of $\gamma_2^2$) and efficiency normalized to the Carnot limit $\eta/\eta_C$ in the weak coupling regime $\kappa=0.001$ as a function of the driving frequency $\Omega$ and the frequency of the peak ${\omega}_1$ (in units of $\omega_0$). Temperatures are set to $T_1=0.2\omega_0$ and $T_2=2\omega_0$, while $\gamma_1=\gamma_2=0.02\omega_0$. The dashed line in panel (b) shows the resonance condition $\omega_1=\omega_0-\Omega$.}
\label{fig:2new}\end{figure}
Such environment can be physically realized with cavity architectures~\cite{aspelmeyer, cottet, scigliuzzo, barzanjeh22} (see the sketch in Fig. \ref{fig:2new}(a)). For instance, in cavity optomechanics~\cite{aspelmeyer} a mechanical oscillator (the QHO of frequency $\omega_0$) is embedded in a optical cavity. Then, a laser detuning is imposed on the bare cavity, in order to adjust the resulting frequency resonance $\omega_1$ close to $\omega_0$, like in sideband-resolved cooling experiments~\cite{sidebandcooling}. A possible way to implement a temporal modulation of the system/bath coupling $g_1(t)$ is to superimpose a modulation (by optical pulse or mechanical vibration) to one of the cavity mirrors of characteristic frequency $\Omega$. For clarity we now introduce the dimensionless parameter $\kappa\equiv d_1/(\omega_0^2\omega_1^2)$, which governs the coupling strength.

\subsection{Performance in the weak coupling regime ($\kappa\ll 1$)}
\label{subsec:perfweak}
In the weak coupling regime ($\kappa \ll 1$), one always finds the possibility for a heat engine. Indeed, under the assumption of a sufficiently sharp peak around $\omega_1$ and looking at Eq.~\eqref{f_def}, one can argue that the dominant contributions are for either $\omega_+ \simeq \omega_1$ or $\omega_- \simeq \omega_1$. In the former case, when one can drop the off--resonance contribution due to $\mathcal{J}_1(\omega_-)$, to have $P<0$ one needs
\begin{equation}
\label{eq:templor1}
n_B\left(\frac{\omega_1}{T_1}\right)>n_B\left(\frac{\omega_1-\Omega}{T_2}\right)\implies T_1>\frac{\omega_1}{\omega_1-\Omega}T_2\,,
\end{equation}
while when $\omega_{-}\approx\omega_1$ with analogous reasonings one obtains the condition
\begin{equation}
\label{eq:templor2}
n_B\left(\frac{\omega_1+\Omega}{T_2}\right)>n_B\left(\frac{\omega_1}{T_1}\right)\implies T_1<\frac{\omega_1}{\omega_1+\Omega}T_2\,.
\end{equation}
Figures~\ref{fig:2new}(b--c) show the corresponding power and the efficiency $\eta\equiv-P/J_1$ for a representative temperature arrangement, $T_1/T_2=0.1$. We note that, having chosen $\gamma_2\ll\omega_0$, one finds $\Im \chi_0(\omega)$ peaked around the QHO frequency $\omega_0$. The behavior of the average power in panel (b) well agrees with the resonance condition $\omega_1 \simeq \omega_0 - \Omega$, along which the maximum power occurs (see the dashed line in the plots)~\cite{note3}. With regard to engine efficiency, using the results shown in Appendix~\ref{appendix6} (see Eqs.~(\ref{app_eq:ppd})--(\ref{app_eq:j2pd})) along this resonance one has that $P\approx P^{(-1)}$ and $J_{\nu}\approx J_{\nu}^{(-1)}$ so that $\eta=-P/J_1\approx-P^{(-1)}/J_1^{(-1)}$ and from the expressions quoted above one immediately finds
\begin{equation*}
\label{eq:efflordown}
\eta=1-\frac{\omega_1}{\omega_1+\Omega}<\eta_C\,,
\end{equation*}
where $\eta_C=1-T_1/T_2$ is the efficiency of the Carnot machine. Indeed, from Eq.~\eqref{eq:templor2} one finds  $\frac{\omega_1}{\omega_1+\Omega}>\frac{T_1}{T_2}$, showing that the efficiency is upper bound by the Carnot limit which can therefore be approached in a realistic parameter window. Note, however, that maximum power and maximum efficiency are reached at different values of $\omega_1$. In particular, inspecting Eq.~(\ref{app_eq:ppd}) one immediately sees that when $\frac{\omega_1}{\omega_1+\Omega}=\frac{T_1}{T_2}$ the power vanishes.\\

Within the model of a cavity of Fig. \ref{fig:2new}(a), one can intuitively interpret the produced power as the unbalance between the energy flux flowing from the contact at temperature $T_2>T_1$ and the energy flux that can be absorbed by the red-shifted cavity ($\omega_1=\omega_0-\Omega$) at the colder temperature $T_1$.\\

Similar results are found for the resonance condition $\omega_+ \simeq \omega_1$ for $T_1 > T_2$. Following steps perfectly analogous to the ones outlined above, one finds the efficiency to be
\begin{equation*}
\eta=1-\frac{\omega_1-\Omega}{\omega_1}<\eta_C\,,
\end{equation*}
where now $\eta_C=1-T_2/T_1$. Also in this case the Carnot limit is achieved when $P\to 0$, when $\frac{\omega_1-\Omega}{\omega_1}=\frac{T_2}{T_1}$.\\

\subsection{Effective model in terms of quantum otto engines}
\label{subsec:effectiveQHO}
The above results can be interpreted in terms of an effective model in which the QHO is regarded as a thermodynamic substance performing a quantum Otto cycle~\cite{Rezek06,Kosloff07}. To illustrate this fact observe that, in the weak coupling regime, the expressions for the power and heat currents are sums over two independent ``channels" labeled by $p=\pm 1$ -- see the last identifications in Eqs.~(\ref{app_eq:ppd})--(\ref{app_eq:j2pd}). For a sharp Lorentzian spectral density, the resonance condition $\omega_1=\omega_0+p\Omega$ implies $\mathcal{J}_1(\omega_0-p\Omega)\ll\mathcal{J}_1(\omega_1)$, which allows to focus only the $p$--th channel ignoring the negligible contribution of the channel $-p$, i.e. to treat the two channels separately.\\

\begin{figure}[t]
\includegraphics[width=8cm,keepaspectratio]{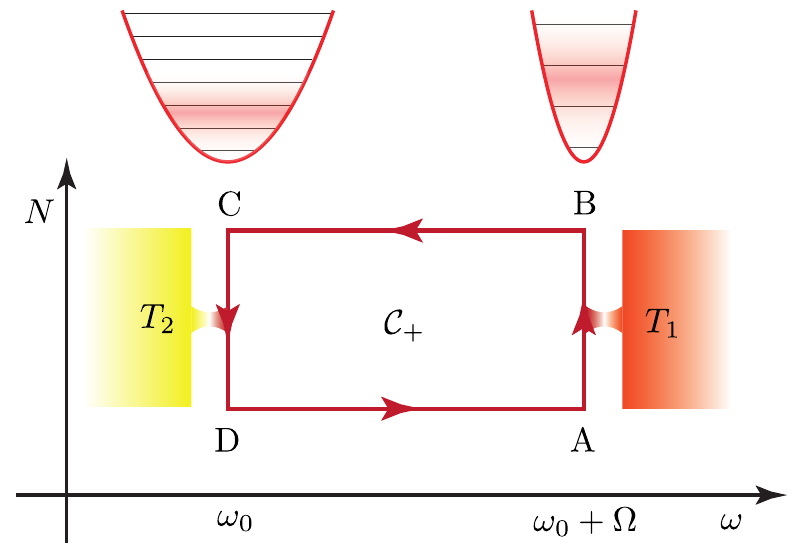}
\caption{Schematic representation of the Otto cycle $\mathcal{C}_+$ in the $N$--$\omega$ plane, where $N$ is the average occupation number of the QHO.}\label{app_fig:ottocycle}	
\end{figure}

Let us consider now the channel $p=+1$, i.e. the resonance $\omega_1=\omega_0+\Omega$ with $T_1>T_2$. We can build an effective model in terms of a Otto engine cycle $\mathcal{C_{+}}$ represented schematically in Fig.~\ref{app_fig:ottocycle} and composed as follows:
\begin{itemize}
    \item an {\em isochoric} transformation A$\to$B along which the QHO is kept at frequency $\omega_0+\Omega$ and allowed to exchange heat with the hot bath at temperature $T_1$ over a characteristic time $\tau_1$;
    \item an {\em isentropic} expansion B$\to$C, where the oscillator is decoupled from the baths and its frequency evolves adiabatically from $\omega_0+\Omega$ to $\omega_0$;
    \item an {\em isochoric} transformation C$\to$D at frequency $\omega_0$ exchanging heat with the cold bath at temperature $T_2$ over a characteristic time $\tau_2$;
    \item an {\em isentropic} compression D$\to$A, where the frequency adiabatically turns back from $\omega_0$ to $\omega_0+\Omega$. .
\end{itemize}
The characteristic time spent along each isentropic branch is $\tau_{\mathrm{is}}>\omega_0^{-1}$, which we assume large enough so that average occupation number of the QHO (denoted here as $N$) is conserved~\cite{Rezek06}: $N_{\mathrm{B}}=N_{\mathrm{C}}$, and $N_{\mathrm{D}}=N_{\mathrm{A}}$. Also, we assume that at the end of each isochor the QHO has thermalized to the corresponding bath: 
$N_{\mathrm{B}}=n_B\left(\frac{\omega_0+\Omega}{T_1}\right)=N_{\mathrm{C}}$ and
$N_{\mathrm{D}}=n_B\left(\frac{\omega_0}{T_2}\right)=N_{\mathrm{A}}$. Thermalization along the isochors takes finite characteristic times $\tau_{1,2}$ which, to lowest order, can be identified with the inverse rates~\cite{breuerbook} $\tau_1\approx\frac{4m\omega_0}{\mathcal{J}_1(\omega_0+\Omega)}$ and $\tau_2\approx\frac{4m\omega_0}{\mathcal{J}_2(\omega_0)}$.\\

The heat exchanged with the contacts are given by~\cite{Rezek06} $\mathcal{Q}_1=(\omega_0+\Omega)(N_{\mathrm{B}}-N_{\mathrm{A}})$ and $\mathcal{Q}_2=\omega_0(N_{\mathrm{D}}-N_{\mathrm{C}})\equiv-\omega_0(N_{\mathrm{B}}-N_{\mathrm{A}})$ while the total work reads $\mathcal{W}=N_{\mathrm{B}}[\omega_0-(\omega_0+\Omega)]+N_{\mathrm{A}}[(\omega_0+\Omega)-\omega_0]\equiv-\Omega(N_{\mathrm{B}}-N_{\mathrm{A}})$. 

To complete the mapping we remind that we deal with a perturbative regime for the spectral density $\mathcal{J}_1$ and thus $\tau_1\gg\tau_2$, and that to obtain Eqs.~(\ref{app_eq:ppd})--(\ref{app_eq:j2pd}) the regime $\gamma_2\ll\omega_0$ has been considered -- see Appendix~\ref{appendix5}, in particular the argument leading to Eq.~(\ref{app_eq:deltas}) -- so that $\tau_2\gg\tau_{\mathrm{is}}$. Thus the total time spent on the cycle is $\approx\tau_{1}$ which allows to estimate the average heat currents as $\mathcal{Q}_{\nu}/\tau_1\equiv J_{\nu}^{(+1)}$ and the average power as $\mathcal{W}/\tau_1\equiv P^{(+1)}$ -- see Eq.~(\ref{app_eq:ppd}).\\

For the cycle to operate as a heat engine one needs $\mathcal{W}<0$ which implies
\begin{equation*}
T_1>\frac{\omega_0+\Omega}{\omega_0}T_2\rightarrow T_1>\frac{\omega_1}{\omega_1-\Omega}T_2\,,    
\end{equation*}
as also found previously. The efficiency $\eta_{+}=\frac{-P^{(+1)}}{J_1^{(+1)}}$ of $\mathcal{C}_+$ is given by
\begin{equation*}
\eta_{+}=1-\frac{\omega_0}{\omega_1}\equiv1-\frac{\omega_1-\Omega}{\omega_1}\,,    
\end{equation*}
is governed by the compression ratio of the QHO and is in accordance with the physics of a Otto cycle and with the results quoted in the previous section. If $T_1<\frac{\omega_0+\Omega}{\omega_0}T_2$ one instead finds $P^{(+1)}>0$.\\

With similar arguments one can interpret the resonance $\omega_1=\omega_0-\Omega$ (channel $p=-1$), where the cycle $\mathcal{C}_-$ is dominant. It is comprised of two isentrops operating between the frequencies $\omega_0$ and $\omega_0-\Omega$ and two isochors where the QHO is kept, with fixed frequency $\omega_0$ (or $\omega_0-\Omega$), in contact with the hot bath at temperature $T_2$ (or the cold bath at temperature $T_1$). Identifying the time spent in contact with the bath at $T_1$ as $\tau\approx\frac{4m\omega_0}{\mathcal{J}_1(\omega_0-\Omega)}$, which is also the longest time in the cycle, allows to evaluate the heat currents and the power with reasonings similar to those made for ${\mathcal{C}}_+$. The expressions again coincide with $J_{\nu}^{(-1)}$ and $P^{(-1)}$ -- see Eqs.~(\ref{app_eq:ppd}),~(\ref{app_eq:j1pd}).\\

For $\mathcal{C}_{-}$ to operate as an engine, one needs 
\begin{equation*}
T_2>\frac{\omega_0}{\omega_0-\Omega}T_1\rightarrow T_1<\frac{\omega_1}{\omega_1+\Omega}T_2\,,    
\end{equation*}
in accordance to what discussed in the previous section. In the heat engine regime, the efficiency $\eta_{-}=\frac{-P^{(-1)}}{J_2^{(-1)}}$ is
\begin{equation*}
\eta_{-}=1-\frac{\omega_1}{\omega_0}\equiv1-\frac{\omega_1}{\omega_1+\Omega}\,.
\end{equation*}
Also this result agrees with the ones reported in the previous section.\\

An important remark must be made: the effective model leading to $\mathcal{C}_-$ breaks down when $\Omega\to\omega_0$: here the effective volume of the QHO and the cycle time diverge and correspondingly $P^{(-1)}\to 0$. When $\Omega>\omega_0$, one can see that $P^{(-1)}>0$ and $J_{\nu}^{(-1)}<0$: Now the $p=-1$ channel acts as a ``heater" absorbing work from the driving mechanism and discharging it into both baths~\cite{Buffoni}.

The analysis conducted above can also be applied to spectral densities other than the Lorentzian one. However, when ${\mathcal{J}}_1(\omega)$  is not sharply peaked the contributions of the two channels $p=\pm 1$ cannot be clearly separated. In this case one can interpret the results as the action of two thermal machines running in parallel: heat currents $J_{\nu}$ through the baths split/recombine into the two channels $J_{\nu}^{(p)}$ and the total power $P$ is the net sum of the powers $P^{(p)}$ exchanged by each machine. When $\Omega<\omega_0$ the two machines perform the Otto cycles $\mathcal{C}_{\pm}$ discussed above, while for $\Omega>\omega_0$ only the channel $p=+1$ behaves as a Otto cycle, while $p=-1$ acts as a heater. From the discussion above it is clear that for given $\Omega$ and temperature ratio $T_2/T_1$ only at most one channel can operate as a heat engine. Therefore, either $\frac{T_2}{T_1}>\frac{\omega_0}{\omega_0-\Omega}$ or $\frac{T_2}{T_1}<\frac{\omega_0}{\omega_0+\Omega}$ is a {\em necessary} but not sufficient condition for obtaining $P<0$ and the precise balance between the power exchanged by the two channels must be studied~\cite{notaFW}.

\subsection{Beyond the weak coupling regime}
\label{subsec:beyond}
\begin{figure}
\includegraphics[width=8cm,keepaspectratio]{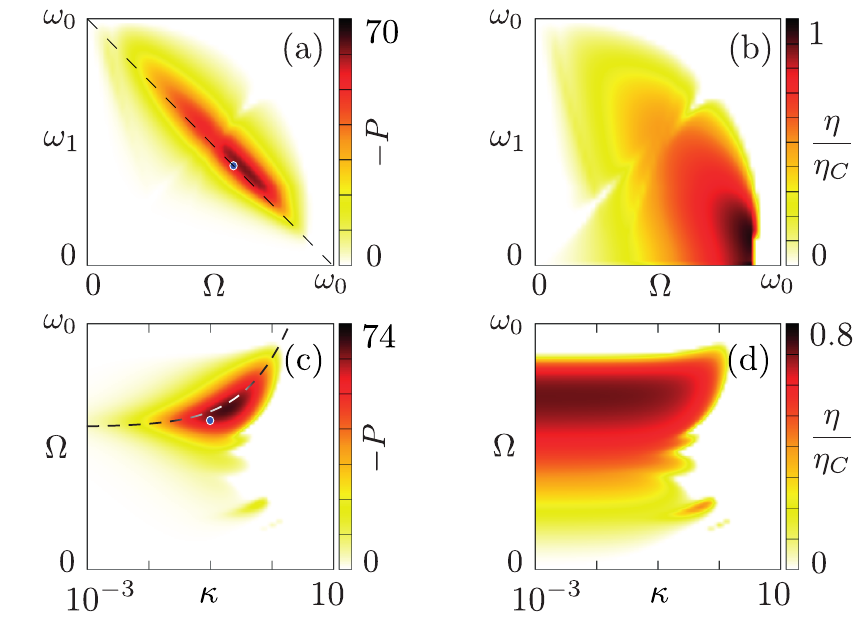}
\caption{non--Markovian heat engine at strong coupling: 
panels (a) and (b) show the average power (in unit of $\gamma_2^2$) and efficiency normalized to the Carnot limit $\eta/\eta_C$ for $\kappa=0.1$ as a function of the driving frequency $\Omega$ and the frequency of the peak ${\omega}_1$.  Panels (c) and (d) show the average power (units $\gamma_2^2$) and efficiency normalized to the Carnot limit as a function of $\kappa$ and $\Omega$ for $\omega_1=0.4\ \omega_0$. The blue point marks corresponding parameters between Panels (a) and (c). Here, $T_1=0.2\omega_0$ and $T_2=2\omega_0$ while $\gamma_1=\gamma_2=0.02\omega_0$.}\label{fig:3new}	
\end{figure}
We conclude this section studying a regime beyond weak coupling. In Figs.~\ref{fig:3new}(a,b) we show heat engine performance, obtained by numerically solving Eq.~(\ref{app_sfloquet1}) and evaluating the expressions for average power in Eq.~(\ref{Paverage}) and corresponding heat currents~\cite{notaPC}. A clear broadening of the power resonance line can be observed (panel (a)). Indeed, the maximum power is no longer achieved along the "bare" resonance $\omega_1=\omega_0-\Omega$ (see the dashed line) but now appears detuned. Also the efficiency, reported in panel (b), displays an overall broadening.

To further inspect the strong-coupling regime we also show in Fig. \ref{fig:3new}(c,d) the average power and efficiency as a function of the coupling parameter $\kappa$ and the driving frequency $\Omega$ near the maximum of Fig. \ref{fig:3new}(a). Operating the engine beyond weak coupling allows to achieve sensibly larger power outputs. Indeed, the maximum power occurs at moderate/strong coupling $\kappa\approx 0.2$ and is over 10 times larger than the maximum power obtained at weak coupling $\kappa=10^{-3}$. However, the average power is a non--monotonic function of $\kappa$ and for $\kappa\gtrsim 1$ the heat engine is lost. Looking at Fig. \ref{fig:3new}(d), the maximum efficiency is achieved operating the engine in the weak coupling regime $\kappa\ll 1$. Therefore, the parameter $\kappa$ can be used to tune the trade-off between power and efficiency. As a final remark, we note that stronger coupling strengths induce a marked detuning of the resonance frequency, due to an energy  renormalization which can be captured by solving self--consistently $\omega^2-\omega_0^2-\mathrm{Re}\left[\tilde{k}_0(\omega)\right]=0$ and $\omega_1=\omega-\Omega$. The solution is shown as a dashed line in Fig. \ref{fig:3new}c).\\

\section{Conclusions and outlook}
\label{sec:conclusions}
We have shown that by properly  modulating the system-bath coupling it is possible to obtain a working heat engine. Here, non--Markovianity is a useful resource for quantum thermodynamics, in that it allows for efficient dynamical heat engines, even approaching Carnot efficiency. Our results open up new possibilities for the exploitation of non--Markovianity. For instance, one could consider the combined effect of  modulating couplings and driving the system, looking for a cooperative effect to enhance  the performance of thermal machines. 
In such a quest, machine learning tools~\cite{simone, plastina, noe, khait2022} could prove to be useful. Regarding possible implementations, our results on structured environment could be tested in the field of cavity optomechanics, which are emerging as an interesting platform for new quantum technologies~\cite{aspelmeyer, barzanjeh22, barzanjeh2, pontin}. We believe that these findings are not restricted to the investigated working medium of a QHO, and an interesting follow-up would be to investigate the role of non--Markovian contributions with different quantum system, for instance one or more qubits, that can be 
integrated in superconducting waveguide quantum electrodynamics architectures~\cite{cottet, scigliuzzo, viennot, lu21, rodrigues}.\\ 

{\it Acknowledgments.}--- F.C. and M.S. acknowledge support by the ``Dipartimento di Eccellenza MIUR 2018-2022''. G.B. acknowledges financial support by the  Julian Schwinger Foundation (Grant JSF-21-04-0001) and by INFN through the project ``QUANTUM''. We thank an anonymous referee for useful suggestions.\\

\appendix
\section{Proof that $\mathcal{A}=0$}
\label{appendix1}
Here we prove that
\begin{equation*}
{\cal A}=\frac{1}{\cal T}\int_t^{t+{\cal T}}\!\!\!\!\!dt'\,{\rm Tr}\Big[
\frac{d}{dt'}\Big(H_{\rm QHO}(t')+H_{\rm int}^{(t')}(t')\Big)\rho(t_0)\Big]=0\,.    
\end{equation*}
A convenient strategy is to rewrite the above quantity as 
\begin{eqnarray}
{\cal A}&=&\frac{1}{\cal T} \Big[\langle H_{\rm int}^{(t+{\cal T})}(t+{\cal T})\rangle - \langle H_{\rm int}^{(t)}(t)\rangle\nonumber\\
&+&\langle H_{\rm QHO}(t+{\cal T})\rangle -
 \langle H_{\rm QHO}(t)\rangle\Big]\,,\nonumber
\end{eqnarray}
Below, we will show that $ \langle H_{\rm int}^{(t)}(t)\rangle$ and $ \langle H_{\rm QHO}(t)\rangle$ are periodic functions with period $\cal T$, which eventually implies $\mathcal{A}=0$. The time dependence of the interaction term is given by
\begin{eqnarray}
H_{\rm int}^{(t)}(t)&=&-x(t) \sum_{\nu=1}^2\sum_{k=1}^{\infty}g_\nu(t)c_{k,\nu}X_{k,\nu}(t) \nonumber\\
&+&x^2(t) \sum_{\nu=1}^2\sum_{k=1}^{\infty} g_{\nu}^2(t)\frac{c^2_{k,\nu}}{2m_{k,\nu} \omega^2_{k,\nu}}.\nonumber
\end{eqnarray}
Using the equations of motion for the position $x(t)$ and momentum $p(t)$ of the QHO
\begin{equation}
\dot{x}(t)=\frac{p(t)}{m}\ \ ;\ \ \dot{p}(t)=m\ddot{x}(t)\,,\nonumber
\end{equation}
with $\ddot{x}(t)$ given in Eq.~(\ref{eq:diffqoper}) and those for the position and momentum of the degrees of freedom of the baths
\begin{eqnarray}
\label{app_eq:EOM1SP}
\dot{X}_{{k,\nu}}(t)&=&\frac{P_{k,\nu}(t)}{m_{k,\nu}}\,,\nonumber\\
\dot{P}_{{k,\nu}}(t)&=&- m_{k,\nu}\omega^2_{k,\nu} X_{k,\nu}(t) +g_\nu(t) c_{k,\nu} x(t)\,\nonumber
\end{eqnarray}
we can rewrite $H_{\rm int}^{(t)}(t)$ in terms of the system position operators $x(t)$ alone, as
\begin{eqnarray}
H_{\rm int}^{(t)}(t)&=&-mx(t)\ddot{x}(t)\nonumber\\
&-&x^2(t)\Big[m\omega_0^2+ \sum_{\nu=1}^2\sum_{k=1}^{\infty} g_{\nu}^2(t)\frac{c^2_{k,\nu}}{2m_{k,\nu} \omega^2_{k,\nu}}\Big]\,.\nonumber
\end{eqnarray}
Similarly, for the QHO term in Eq.~(\ref{eq:HQHO}) we have 
\begin{equation}
H_{{\rm QHO}}(t)= \frac{m}{2}\left[\dot{x}^2(t)+ \omega_0^2x^2(t)\right].\nonumber
\end{equation}
The above expressions show that their quantum ensemble averages can be written as correlators of the QHO position operator {\em only}. Indeed we have
\begin{eqnarray}
&&\langle H_{\rm int}^{(t)}(t)\rangle=-m M_{x\ddot x}(t)\nonumber\\
&&-\Big[m\omega_0^2+ \sum_{\nu=1}^2\sum_{k=1}^{\infty} g_{\nu}^2(t)\frac{c^2_{k,\nu}}{2m_{k,\nu} \omega^2_{k,\nu}}\Big]M_{xx}(t)\,,\nonumber\\
&&\langle H_{{\rm QHO}}(t)\rangle = \frac{m}{2}M_{\dot x\dot x}(t) + \frac{m\omega_0^2}{2}M_{xx}(t)\,,\nonumber
\end{eqnarray}
where
\begin{equation*}
M_{xx}(t)={\rm Tr}\Big[x(t)x(t)\rho(t_0)\Big]\,,
\end{equation*}
\begin{equation*}
M_{x\ddot x}(t)={\rm Tr}\Big[x(t)\ddot{x}(t)\rho(t_0)\Big]\,, 
\end{equation*}
\begin{equation*}
M_{\dot x\dot x}(t)={\rm Tr}\Big[\dot{x}(t)\dot{x}(t)\rho(t_0)\Big]\,.
\end{equation*}
Note that all these correlators can be represented in terms of the function
\begin{equation}
\label{app_averagemSP}
M(t,s)={\rm Tr}\Big[x(t)x(s)\rho(t_0)\Big]
\end{equation}
as:
\begin{eqnarray}
M_{xx}(t)&=&\lim_{s\to t} M(t,s)\,,\label{eq:der1}\\
M_{x\ddot x}(t)&=&\lim_{s\to t}\frac{d^2}{ds^2} M(t,s)\,,\label{eq:der2}\\
M_{\dot x\dot x}(t)&=&\lim_{s\to t}
\frac{d}{dt}\frac{d}{ds}M(t,s)\,.\label{eq:der3}
\end{eqnarray}
We then focus on the evaluation of the quantum average of $M(t,s)$. Plugging into Eq.~(\ref{app_averagemSP}) the time evolution given in Eq.~(\ref{app_solutionSP}) we obtain 
\begin{eqnarray}
M(t,s)&=&\frac{1}{m^2}\int_{-\infty}^{+\infty}\!\!\!\!\!\mathrm{d}t_1 \int_{-\infty}^{+\infty}\!\!\!\!\!\mathrm{d}t_2
G(t,t_1) G(s,t_2)\nonumber\\
&\times&\sum_{\nu=1}^{2}\sum_{\nu'=1}^{2}g_\nu(t_1)g_{\nu'}(t_2)\langle\xi_\nu(t_1)\xi_{\nu'}(t_2)\rangle=\nonumber\\
&=&
\frac{1}{m^2}\int_{-\infty}^{+\infty}\!\!\!\!\!\mathrm{d}t_1 \int_{-\infty}^{+\infty}\!\!\!\!\!\mathrm{d}t_2
G(t,t_1) G(s,t_2)\nonumber\\
&\times&\sum_{\nu=1}^{2}g_\nu(t_1)g_\nu(t_2){\cal L}_\nu(t_1-t_2)\,,\label{app_solutionMSP}
\end{eqnarray}
where in the second equality we have inserted the bath correlator $\langle \xi_\nu (t)\xi_{\nu'}(t')\rangle =\delta_{\nu,\nu'}{\cal L}_\nu (t-t')$ with ${\cal L}_\nu (t-t')$ given in Eq.~(\ref{correlator}).
The time integrals are performed using the Fourier representations in Eq.~(\ref{app_transform}) for the  Green functions $G(t,t_1)$ and $G(s,t_2)$, Eq.~(\ref{app_seriesegSPr}) for the driving $g_\nu(t_1)$ and 
$g_{\nu}(t_2)$, and  
\begin{equation*}
{\cal L}_\nu(t_1-t_2)=\int_{-\infty}^{+\infty}\!\!\frac{\mathrm{d}\omega}{2\pi}e^{-i\omega(t_1-t_2)}{\tilde {\cal L}}_\nu(\omega)
\end{equation*}
for the bath correlator. We finally obtain
\begin{eqnarray}
M(t,s)&=&\!\!\!\!\!\!\!\!\!\!\!\!\!\!\sum_{n_1,n_2,\mu_1\mu_2=-\infty}^{+\infty}\sum_{\nu=1}^{2}g_{n_1,\nu}g_{n_2,\nu}\int_{-\infty}^{+\infty}\frac{\mathrm{d}\omega}{2\pi m^2}
{\tilde {\cal L}}_\nu(\omega)\nonumber\\
&\times&\tilde{G}_{\mu_2}(-\omega+n_2\Omega)\tilde{ G}_{\mu_1}(\omega+n_1\Omega)\nonumber\\
&\times&e^{-it(\omega+(n_1+\mu_1)\Omega)}e^{is(\omega-(n_2+\mu_2)\Omega)}\,.\nonumber
\end{eqnarray}
Notice that now the times $t$ and $s$ only appear in the exponential factors: this means that   after derivatives with respect to $t$ and $s$ as appropriate according to the definitions in Eqs.~(\ref{eq:der1})--(\ref{eq:der3}), the limit $s\to t$ implies a time dependent term always of the form $e^{-i\Omega t(n_1+n_2+\mu_1+\mu_2)}$. This shows a clear periodicity with respect to the cycle time $\cal T$. This result demonstrates that the required correlators are indeed periodic:  $M_{xx}(t+{\cal T})=M_{xx}(t)$, $M_{x\ddot x}(t+{\cal T})=M_{x\ddot x}(t)$ and 
$M_{\dot x\dot x}(t+{\cal T})=M_{\dot x\dot x}(t)$. Since, as shown above, these correlators are the building blocks of $\langle H_{\rm int}^{(t)}(t)\rangle$ and $\langle H_{{\rm QHO}}(t)\rangle $ we eventually arrive to the conclusion that {\em the quantum averages of the interaction term and of the QHO part are periodic}. This finally proves the key result that ${\cal A}=0$ and then that the cycling average of the total  power  is totally balanced {\em only} by the reservoir heat currents -- see Eq.~(\ref{app_firstlaw1}).

\section{Average power and heat currents for a monochromatic drive}
\label{appendix2}
Here we derive a compact form for the power and heat currents by focusing on the case discussed in the main part, namely a constant coupling to the bath $\nu=2$ and a monochromatically modulated coupling to the bath $\nu=1$, see Eq.~(\ref{eq:drivingscheme}). It is useful to begin recalling the general expressions of the average total power $P$ and  heat currents $J_{\nu}$ performed using Eqs.~(\ref{app_PSP}) and~(\ref{app_JSP}) as explained in Ref.~\cite{carrega_prxquantum}. We have:
\begin{widetext}
\beq
P&=&\Omega\!\!\!\!\sum_{n_1,n_2=-\infty}^{+\infty}\sum_{\nu=1}^2 n_1g_{n_1,\nu}g_{n_2,\nu}\int_{-\infty}^{+\infty}\frac{\mathrm{d}\omega}{2\pi m}\Big\{i{\cal J}_\nu(\omega) \coth(\frac{\omega}{2T_\nu}) \tilde{ G}_{-(n_1+n_2)}(\omega+n_2\Omega)\nonumber\\
&&\!\!\!\!\!\!\!\!\!\!\!\!+\sum_{\nu_1=1}^{2} \sum_{\mu=-\infty}^{+\infty}\sum_{n_3, n_4=-\infty}^{+\infty}\!g_{n_3,\nu_1}g_{n_4,\nu_1}\!\frac{{\cal J}_{\nu_1}(\omega)}{m}\coth(\frac{\omega}{2T_{\nu_1}}){\cal J}_{\nu}(\omega-\Omega(n_2+n_4+\mu))\tilde{ G}_{\mu}(-\omega+n_4\Omega)\tilde{ G}_{-(n_{\rm tot}+\mu)}(\omega+n_3\Omega)\Big\}\,,\qquad\label{app_Paverage}
\eeq
\beq
\label{app_javerage}
J_\nu&=&\!\!\!\!\sum_{n_1,n_2=-\infty}^{+\infty}
g_{n_1,\nu}g_{n_2,\nu}\int_{-\infty}^{+\infty}\frac{\mathrm{d}\omega}{2\pi m}\Big\{-i {\cal J}_\nu(\omega)\omega\coth(\frac{\omega}{2T_\nu}) \tilde{ G}_{-(n_1+n_2)}(\omega+n_2\Omega)-\sum_{\nu_1=1}^{2} \sum_{\mu=-\infty}^{+\infty}
\sum_{n_3, n_4=-\infty}^{+\infty}\!g_{n_3,\nu_1}g_{n_4,\nu_1}\nonumber\\
&\times&\frac{{\cal J}_{\nu_1}(\omega)}{m}\coth(\frac{\omega}{2T_{\nu_1}})[\omega-\Omega(n_2+n_4+\mu)]{\cal J}_{\nu}(\omega-\Omega(n_2+n_4+\mu))\tilde{ G}_{m_1}(-\omega+n_4\Omega)\tilde{ G}_{-(n_{\rm tot}+\mu)}(\omega+n_3\Omega)\Big\},
\eeq
\end{widetext}
where $n_{\rm tot}=n_1+n_2+n_3+n_4$. We remind that the spectral densities ${\cal J}_{\nu}(\omega)$ are odd functions of frequency. We now specialize to the driving considered in this work, with Fourier coefficients $g_{n,1}= (\delta_{n,1}+\delta_{n,-1})/2$ and $g_{n,2}=\delta_{n,0}$. Then the kernel in  Eq.~\eqref{app_chikappaSP} reduces to
\begin{eqnarray}
\tilde{k}_0(\omega)&=&-i\omega\tilde{\gamma}_2(\omega)+\!\!\!\sum_{n=\pm2}\!\!\!\tilde{k}_{n}(\omega)\,,\label{app_s1kappan1}\\
\tilde{k}_{\pm2}(\omega)&=&-\frac{i}{4} \omega_{\pm}\tilde{\gamma}_1 (\omega_{\pm})\label{app_s1kappan2}
\end{eqnarray}
with $\omega_{\pm}=\omega\pm\Omega$. This shows that  only the kernels $\tilde{k}_{0,\pm 2}(\omega)$ are different from zero.
We now plug the expressions of $g_{0,\nu}$ into Eq.~\eqref{app_Paverage} to write down the average power. Notice that only the $\nu=1$ term contributes to $P$, that can be conveniently decomposed into two contributions $P=P^{(a)} + P^{(b)}$. The first, stemming from the first line of Eq.~\eqref{app_Paverage} reads
\begin{eqnarray}
P^{(a)}&=&-\frac{\Omega}{4\pi m}\int_{-\infty}^{+\infty}\!\!\mathrm{d}\omega{\cal J}_1(\omega)\coth(\frac{\omega}{2T_1})\nonumber\\
&\times&\Im\left[\tilde{ G}_{0}(\omega-\Omega)-\tilde{ G}_{2}(\omega-\Omega)\right]\,,\label{app_p1a}
\end{eqnarray}
while the second originates from the second line of the same equation and it is given by
\begin{widetext}
\beq
\label{app_p1b}
P^{(b)}&=&\Omega\sum_{n_1,n_2=-\infty}^{+\infty}
n_1g_{n_1,1}g_{n_2,1}\sum_{n_3, n_4=-\infty}^{+\infty}\sum_{\nu_1=1}^{2} g_{n_3,\nu_1}g_{n_4,\nu_1}\sum_{\mu=-\infty}^{+\infty}\int_{-\infty}^{+\infty}\!\!\frac{\mathrm{d}\omega}{2\pi m^2}{\cal J}_{\nu_1}(\omega)
\coth(\frac{\omega}{2T_{\nu_1}})\nonumber\\&&\nonumber\\
&&{\times} {\cal J}_{1}(\omega-\Omega(n_2+n_4+\mu))\tilde{ G}_{\mu}(-\omega+n_4\Omega)\tilde{ G}_{-(n_{\rm tot}+\mu)}(\omega+n_3\Omega).
\eeq
\end{widetext}
To obtain Eq.~\eqref{app_p1a}, we have used that $n_1,n_2=\pm1$, and the property of the Floquet coefficients
\be
\label{app_property1}
\tilde{G}_\mu(\omega)=\tilde{G}_{-\mu}^* (-\omega)~.
\ee
Exploiting the symmetry property (see Appendix~\ref{appendix3} for details) 
\be
\label{app_property2}
\tilde{ G}_{\mu}\left(\omega-\frac{\mu}{2}\Omega\right)=\tilde{ G}_{-\mu}\left(\omega+\frac{\mu}{2}\Omega\right)
\ee
one sees that the last term in the square brackets of Eq.~\eqref{app_p1a} has a null contribution upon integration giving then
\begin{eqnarray}
P^{(a)}&=&-\frac{\Omega}{4\pi m}\int_{-\infty}^{+\infty}\!\!\mathrm{d}\omega{\cal J}_1(\omega)\coth(\frac{\omega}{2T_1})\nonumber\\
&\times&\Im\left[\tilde{ G}_{0}(\omega-\Omega)\right]\,.\label{app_p1afinal}
\end{eqnarray}
Using Eq.~\eqref{app_property1} and renaming $\mu+n_2+n_4\to \mu$ we can then rewrite the $P^{(b)}$ term as
\begin{widetext}
\beq
P^{(b)}&=&\frac{\Omega}{2\pi m^2}\!\!\!\!\sum_{n_1,n_2=-\infty}^{+\infty}\!\!\!\!\!\!
n_1g_{n_1,1}g_{n_2,1}\!\!\!\!\sum_{n_3, n_4=-\infty}^{+\infty}\sum_{\nu_1=1}^{2} \!g_{n_3,\nu_1}g_{n_4,\nu_1}\sum_{\mu=-\infty}^{+\infty}\int_{-\infty}^{+\infty}\!\!\mathrm{d}\omega{\cal J}_{\nu_1}(\omega)
\coth(\frac{\omega}{2T_{\nu_1}})\nonumber\\
&&\times{\cal J}_{1}(\omega-\mu\Omega)\Re\left[\tilde{ G}_{\mu-(n_2+n_4)}(-\omega+n_4\Omega)\tilde{ G}_{-(\mu+n_1+n_3)}(\omega+n_3\Omega)\right]\label{app_P1bbis}
.
\eeq
\end{widetext}
We now insert the explicit form of $g_{n,\nu}$, and for notational convenience we write $P^{(b)}=P^{(b,1)}+P^{(b,2)}$ corresponding to the $\nu_1=1,2$ terms in the above expression. The former contribution reads
\begin{eqnarray}
P^{(b,1)}&=&\frac{\Omega}{32\pi m^2}\int_{-\infty}^{+\infty}\!\!\!\!\mathrm{d}\omega{\cal J}_{1}(\omega)\coth(\frac{\omega}{2T_{1}})\nonumber\\
&&\!\!\!\!\!\!\!\!\!\!\!\!\!\!\!\!\!\!\!\!\!\!\!\!\!\times\sum_{\mu=-\infty}^{+\infty}{\cal J}_{1}(\omega-\mu\Omega)\Re\Big[\!\!\!\!\!\sum_{n_1,n_3=\pm 1}\!\!\!\!
n_1\tilde{ G}_{-(\mu+n_1+n_3)}(\omega+n_3\Omega)\nonumber\\
&&\!\!\!\!\!\!\!\!\!\!\!\!\!\!\!\!\!\!\!\!\!\!\!\!\!\times\sum_{n_2, n_4=\pm 1}\tilde{ G}_{\mu-(n_2+n_4)}(-\omega+n_4\Omega)\Big]\,.\label{app_P1b1}
\end{eqnarray}
Taking the real part of the sum in the last square bracket and using again Eq.~\eqref{app_property1}, one has
\begin{eqnarray}
\!\!\!\!\!\!\!\!\!&&\!\!\!\!\!\!\!\!\!P^{(b,1)}=\frac{\Omega}{16\pi m^2}\int_{-\infty}^{+\infty}\!\!\!\mathrm{d}\omega{\cal J}_{1}(\omega)\coth(\frac{\omega}{2T_{1}})\nonumber\\
\!\!\!\!\!\!\!\!\!\!\!&&\!\!\!\!\!\!\!\!\!\!\!\times\!\!\!\!\sum_{\mu=-\infty}^{+\infty}\!\!\!\!{\cal J}_{1}(\omega-\mu\Omega)
|\tilde{ G}_{\mu}(-\omega+\Omega)+\tilde{ G}_{\mu+2}(-\omega-\Omega)|^2.\label{app_P1b1final}
\end{eqnarray}
Note that since $\mathcal{J}_1(\omega)\coth(\frac{\omega}{2T_{1}})>0$ and since the last factor in the summand is positive, the sign of $P^{(b,1)}$ is governed by the behavior of $\mathcal{J}_1(\omega-\mu\Omega)$ only.
We now consider $P^{(b,2)}$ related to the $\nu_1=2$ contribution in Eq.~\eqref{app_p1b}, where $n_1,n_2=\pm1$ and $n_3=n_4=0$:
\begin{eqnarray}
&&P^{(b,2)}=\frac{\Omega}{8\pi m^2}\int_{-\infty}^{+\infty}\!\!\!\mathrm{d}\omega{\cal J}_{2}(\omega)\coth(\frac{\omega}{2T_{2}})\nonumber\\
&&\Re\Big[\!\!\!\sum_{n_1,n_2=\pm 1}\!\!\!\!\!\!
n_1\tilde{ G}_{\mu-n_2}(-\omega)\tilde{ G}_{-(\mu+n_1)}(\omega)\Big]\,.\label{app_P1b2}
\end{eqnarray}
Performing the sum over $n_1,n_2$, we arrive at
\begin{eqnarray}
&&P^{(b,2)}=\frac{\Omega}{8\pi m^2}\int_{-\infty}^{+\infty}\!\!\!\mathrm{d}\omega{\cal J}_{2}(\omega)\coth(\frac{\omega}{2T_{2}})\nonumber\\
&&\!\!\!\!\sum_{\mu=-\infty}^{+\infty}\sum_{p=\pm 1}\!\!\!|\tilde{ G}_{\mu}(\omega)|^2p{\cal J}_{1}[\omega+(\mu+p)\Omega].\label{app_P1b2final}
\end{eqnarray}
Also here, since $\mathcal{J}_2(\omega)\coth(\frac{\omega}{2T_{2}})>0$ and since the first factor in the summand is positive, the sign of $P^{(b,2)}$ is determined by the terms ${\cal J}_{1}[\omega+(\mu\pm1)\Omega]$. Finally, we sum the three contributions in Eqs.~\eqref{app_p1afinal},~\eqref{app_P1b1final},~\eqref{app_P1b2final} to obtain the expression for the average total power $P=P^{(a)}+P^{(b,1)}+P^{(b,2)}$ as reported in Eq.~\eqref{Paverage}.\\

We conclude this part analyzing the average heat current $J_2$ in Eq.~(\ref{app_javerage}). We recall that the average heat current $J_1$, associated to the reservoir $\nu=1$ isobtained from the  energy conservation relation $J_1=-(P+J_2)$. As above, we separate $J_2=J_2^{(a)}+J_2^{(b)}$ into two contributions, where
\be
\label{app_J2a}
\!\!\!\!J_2^{(a)}=\!\!\int_{-\infty}^{+\infty}\frac{\mathrm{d}\omega}{2\pi m}\ \omega{\cal J}_2(\omega)\coth(\frac{\omega}{2T_2})
\Im\left[\tilde{ G}_{0}(\omega)\right]
\ee
and 
\begin{widetext}
\beq
J_2^{(b)}&=&-\sum_{n_1,n_2=-\infty}^{\infty}\! g_{n_1,2}g_{n_2,2}\!\!\!\sum_{n_3, n_4=-\infty}^{+\infty}\sum_{\nu_1=1}^{2} \!g_{n_3,\nu_1}g_{n_4,\nu_1}\!\sum_{\mu =-\infty}^{+\infty}\int_{-\infty}^{+\infty}\!\!\frac{\mathrm{d}\omega}{2\pi m^2}{\cal J}_{\nu_1}(\omega) \coth(\frac{\omega}{2T_{\nu_1}})[ \omega-\Omega(n_2+n_4+\mu )]\nonumber \\
&{\times}&{\cal J}_{2}(\omega-\Omega(n_2+n_4+\mu ))\tilde{G}_{\mu}(-\omega+n_4\Omega)  \tilde{ G}_{-(n_{\rm tot}+\mu )}(\omega+n_3\Omega).\label{app_J2b}
\eeq
\end{widetext}
Again, using Eq.~\eqref{app_property1}, and letting $\mu+n_2+n_4\to \mu$, we can rewrite
\pagebreak
\beq
J_2^{(b)}&=&-\frac{1}{2\pi m^2}\!\!\!\!\sum_{n_1,n_2=-\infty}^{\infty}\!\!\!\!\!\!
g_{n_1,2}g_{n_2,2}\!\!\!\!\sum_{n_3, n_4=-\infty}^{+\infty}\sum_{\nu_1=1}^{2}\!g_{n_3,\nu_1}g_{n_4,\nu_1}\nonumber\\
&\times&\sum_{\mu=-\infty}^{+\infty}\int_{-\infty}^{+\infty}\!\!\mathrm{d}\omega{\cal J}_{\nu_1}(\omega)
\coth(\frac{\omega}{2T_{\nu_1}})(\omega-\mu\Omega)\nonumber\\
&\times&{\cal J}_{2}(\omega-\mu\Omega)\Re\left[\tilde{G}_{\mu-(n_2+n_4)}(-\omega+n_4\Omega)\right.\nonumber\\
&\times&\left.\tilde{G}_{-(\mu+n_1+n_3)}(\omega+n_3\Omega)\right]\,.\label{app_J2bbis}
\eeq
Now, recalling that $n_1=n_2=0$ for $\nu=2$, performing the sum over $n_3,n_4=\pm1$ in the $\nu_1=1$ term, and collecting $J_2=J_2^{(a)}+J_2^{(b)}$ we arrive at Eq.~\eqref{app_sjaaverage}.

\section{A useful property of the Floquet coefficients}
\label{appendix3}
Here we prove that, for the dynamical couplings considered in this work,
\be
\tilde{ G}_{\mu}\left(\omega-\frac{\mu}{2}\Omega\right)=\tilde{ G}_{-\mu}\left(\omega+\frac{\mu}{2}\Omega\right)\,.\nonumber
\ee
We begin by recalling the expression in Eq.~\eqref{app_s1kappan2} for $\tilde{k}_{\pm 2}(\omega)$. In addition, note that for a generic bath on contact 1 with $\tilde{\gamma}_1(\omega)=\gamma_1 \phi(\omega)$, Eq.~(\ref{app_sfloquet1}) can be conveniently rewritten as
 \be
\!\!\!\!\!\!\tilde{G}_\mu (\omega)\!\!=\!\!D_0(\omega)\delta_{\mu,0}\!\!+\!\!\lambda D_{\mu}(\omega)\!\!\sum_{p=\pm 2}\!\!J_{\mu,\mu-p}(\omega)\tilde{G}_{\mu-p}(\omega)
\, ,\label{app_eq:recur}
 \ee
where we have introduced
\begin{eqnarray}
&&\lambda=\frac{\gamma_1}{\omega_0};\quad D_\mu(\omega)=\chi(\omega+\mu\Omega);\nonumber\\
&&J_{\mu,\mu'}(\omega)=\frac{i\omega_0}{4}\left(\omega+\frac{\mu+\mu'}{2}\Omega\right)\phi\left(\omega+\frac{\mu+\mu'}{2}\Omega\right)\,.\nonumber
\end{eqnarray}
We now write a formal series expansion of $\tilde{G}_\mu(\omega)$ in powers of $\lambda$:
\be
\tilde{G}_\mu(\omega)=\sum_{n\geq 0}\lambda^{n}\tilde{G}_{\mu}^{(n)}(\omega),\label{app_eq:series}
\ee
which is plugged into Eq.~(\ref{app_eq:recur}). Matching order-by-order in $\lambda$ a hierarchy of nested equations for the $n$--th contribution $\tilde{G}_\mu^{(n)}(\omega)$ is obtained. In particular one immediately sees that $\tilde{G}_0^{(0)}(\omega)=D_0(\omega)$ and that
\be
 \tilde{G}_\mu^{(n+1)}(\omega)= D_{\mu}(\omega)\sum_{p=\pm 2}J_{\mu,\mu-p}(\omega)\tilde{G}_{\mu-p}^{(n)}(\omega)\, .\label{app_eq:recur2}
 \ee
From Eq.~(\ref{app_eq:recur2}) one can conclude that:\\

\noindent (1) $\tilde{G}_{2\mu+1}^{(n)}(\omega)\equiv 0$ for all $\mu$ and $n\geq 0$;\\
(2) For given $n\geq 0$ the only possible nonzero $\tilde{G}_\mu^{(n)}(\omega)$ occur for $|\mu|\leq 2n$ with $\mu\in\{-2n,-2n+4,\ldots,2n-4,2n\}$.\\

\begin{figure}
\includegraphics[width=8cm,keepaspectratio]{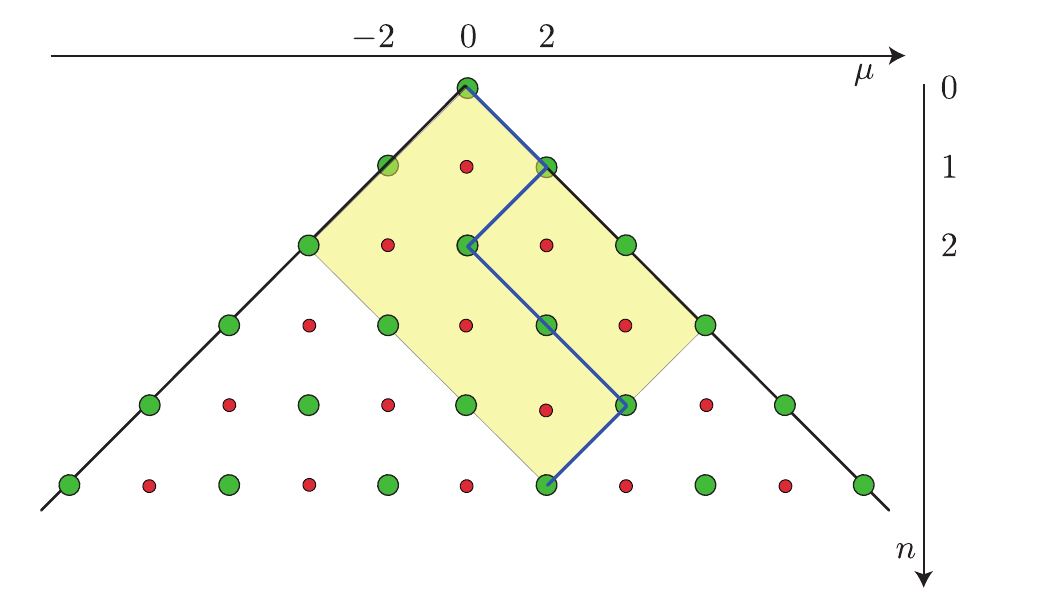}
\caption{The Pascal triangle of dots representing the set of all points associated with $\tilde{G}_{\mu}^{(n)}(\omega)$: Green (red) dots represent a non--zero (zero) $\tilde{G}_{\mu}^{(n)}(\omega)$. The blue line represents one of the paths contributing to $\tilde{G}_{2}^{(5)}(\omega)$, the yellow shaded line represent the region which contains all possible distinct paths contributing to $\tilde{G}_{2}^{(5)}(\omega)$.}\label{app_fig:f1s}	
\end{figure}

One can picture the set of $\tilde{G}_{\mu}^{(n)}(\omega)$ satisfying $n\geq 0$ and $|\mu|\leq2n$ as a lattice of dots on a Pascal triangle, whose rows are labeled by $n$ and whose columns are labeled by $\mu$. This is represented in Fig.~\ref{app_fig:f1s}, where green (red) dots represent the non--zero (zero) $\tilde{G}_{\mu}^{(n)}(\omega)$.\\ 
\noindent Equations~\eqref{app_eq:recur2} can be solved recursively. As an example, to first order one immediately finds $G_{\pm2}^{(1)}(\omega)=D_{\pm 2}(\omega)J_{\pm 2,0}(\omega)D_0(\omega)$. As another example, a non--trivial solution for the second order is 
\begin{eqnarray}
G_{0}^{(2)}(\omega)&=&D_0(\omega)J_{0,2}(\omega)D_{2}(\omega)J_{2,0}(\omega)D_0(\omega)\nonumber\\
\!\!\!\!\!\!\!\!\!\!\!\!&\!\!\!\!\!\!\!\!\!\!\!\!\!\!\!\!+&\!\!\!\!\!\!\!\!D_0(\omega)J_{0,-2}(\omega)D_{-2}(\omega)J_{-2,0}(\omega)D_0(\omega)\,.\label{app_eq:sol02} 
\end{eqnarray}
Each term in the above equation can be interpreted as a {\em path} on the Pascal triangle, linking dots $(\mu,n)$ and $(\mu,n')$ {\em within the triangle} according to the simple rule $|\mu-\mu'|=2$ and $|n-n'|=1$. To each dot a factor $D_{\mu}(\omega)$ is associated, to each link between dots a factor $J_{\mu,\mu'}(\omega)$ is associated.
Explicitly, the two paths representing the terms in Eq.~\eqref{app_eq:sol02} are $(0,2)\to(2,1)\to(0,0)$ and $(0,2)\to(-2,1)\to(0,0)$ respectively. Notice that the index $n$ to the l.h.s. of  gives the number $n$ of links between the $n+1$ dots.\\

\noindent Proceeding with the recursion one quickly realizes that the situation depicted above is general. Indeed, the term $\tilde{G}_{\mu}^{(n)}(\omega)$ consists of a sum of $\mathcal{N}(n,{\mu})=\binom{n}{\frac{2n-{\mu}}{4}}$ terms     
\be
\tilde{G}_{\mu}^{(n)}(\omega)=\sum_{j=1}^{\mathcal{N}(n,\mu)}\tilde{G}_{\mu}^{(n,j)}(\omega)\, ,\label{app_eq:pertsum}
\ee
where $\tilde{G}_{\mu}^{(n,j)}(\omega)$ is associated to one of {\em all} the $\mathcal{N}(n,{\mu})$ distinct paths $\mathcal{P}_j(\mu,n)$ (with $1\leq j\leq\mathcal{N}(n,\mu)$) that connect the dot $(\mu,n)$ with $(0,0)$ with $n$ links that follow the rules $|\mu-\mu'|=2$ and $|n-n'|=1$ as stated above.

More formally, each path $\mathcal{P}_j(\mu,n)$ associated to $\tilde{G}_{\mu}^{(n,j)}(\omega)$ can be represented by the sequence of dots
\be
\mathcal{P}_j({\mu},n)=\{(\mu_{n}^{(j)},n),(\mu_{n-1}^{(j)},n-1),\ldots,(\mu_{0}^{(j)},0)\}\nonumber
\ee
with ``fixed boundaries" $\mu_{n}^{(j)}\equiv\mu$ and $\mu_{0}^{(j)}=0$. One of such path for $\mu=2$ and $n=5$ is shown as a blue line in Fig.~\ref{app_fig:f1s}.\\
\noindent Then, to construct $\tilde{G}_{\mu}^{(n,j)}(\omega)$ we associate a term $D_{\mu_{\nu}}(\omega)$ to each dot in the path and a term $J_{\mu_{\nu},\mu_{\nu+1}}(\omega)$ to each link between consecutive dots (with $0\leq\nu<n$) which leads to
\be
\tilde{G}_{{\mu}}^{(n,j)}(\omega)=D_{\mu_0^{(j)}}(\omega)\prod_{\nu=0}^{n-1}J_{\mu_{\nu}^{(j)},\mu_{\nu+1}^{(j)}}(\omega)D_{\mu_{\nu+1}^{(j)}}(\omega)\,.\label{app_eq:oneterm}
\ee 
The set of all paths $\mathcal{P}(\bar{\mu},n)=\cup_{j=1}^{\mathcal{N}(n,\bar{\mu})}\mathcal{P}_j$ contributing to Eq.~(\ref{app_eq:pertsum}) lies within a rectangular region of the Pascal triangle (a representative example for the case ${\mu}=2$ and $n=5$  is shown as the yellow region in Fig.~\ref{app_fig:f1s}). 
It is simple to see that such rectangle has vertices
\begin{eqnarray}
&&A=(0,0);\quad B=\left(\frac{\mu+2n}{2},\frac{2n+\mu}{4}\right);\nonumber\\
&&C=\left(\frac{\mu-2n}{2},\frac{2n-\mu}{4}\right);\quad D=(\mu,n)\, .\nonumber
\end{eqnarray}

\noindent To prove Eq.~(\ref{app_property2}) for the $n$-th order term we need to shift the argument of $\tilde{G}_{\mu}^{(n)}(\omega)$. To this end, it is useful to observe that $D_\mu(\omega+k\Omega)=D_{\mu+k}(\omega)$ and $J_{\mu,\mu'}(\omega+k\Omega)=J_{\mu+k,\mu'+k}(\omega)$, with $k$ an integer. Geometrically, this means that shifting the argument of $\tilde{G}_{\mu}^{(n)}(\omega)$ by $k\Omega$ is equivalent to shift all paths~\cite{note_paths} that contribute to it (and hence the whole set $\mathcal{P}(\bar{\mu},n)$) by $|k|\Omega$ to the right or to the left according to $\mathrm{Sgn}(k)$.\\

\noindent According to what discussed above, let us denote with $\mathcal{P}_+$ the region containing all the paths contributing to $\tilde{G}_{\mu}^{(n)}\left(\omega-\frac{\mu}{2}\Omega\right)$, with vertices
\begin{eqnarray}
&&A_{+}=\left(-\frac{\mu}{2},0\right);\quad B_{+}=\left(n,\frac{2n+\mu}{4}\right);\nonumber\\ &&C_{+}=\left(-n,\frac{2n-\mu}{4}\right);\quad
D_{+}=\left(\frac{\mu}{2},n\right)\,,\nonumber 
\end{eqnarray}
while the paths contributing to $\tilde{G}_{-2\mu}^{(n)}\left(\omega+\mu\Omega\right)$ belong to the region $\mathcal{P}_{-}$ with vertices
\begin{eqnarray}
&&A_{-}=\left(\frac{\mu}{2},0\right);\quad B_{-}=\left(n,\frac{2n-\mu}{4}\right);\nonumber\\
&&C_{-}=\left(-n,\frac{2n+\mu}{4}\right);\quad
D_{-}=\left(-\frac{\mu}{2},n\right)\,.\nonumber
\end{eqnarray}
Observe that all the factors in Eq.~(\ref{app_eq:oneterm}) actually depend only on the ordered set $\{\mu_{\nu}^{(j)}\}$ but are invariant under any permutation of the second coordinate of each lattice point. This allows to re-order the vertices of $\mathcal{P}_{-}$ in decreasing order of their second coordinate as 
\begin{eqnarray}
&&A_{-}=\left(\frac{\mu}{2},n\right)\equiv D_{+};\quad B_{-}=\left(n,\frac{2n+\mu}{4}\right)\equiv B_{+};\nonumber\\ &&C_{-}=\left(-n,\frac{2n-\mu}{4}\right)\equiv C_{+};\quad
D_{-}=\left(-\frac{\mu}{2},0\right)\equiv A_{+}\,.\nonumber
\end{eqnarray}
This allows us to conclude that the two regions are actually identical. Since shifting the argument only amounts to a rigid translation of the paths and the actual re-ordering performed above corresponds to reading each term from right to left rather than left to right, it follows that to each path of $\tilde{G}_{\mu}^{(n)}\left(\omega-\frac{\mu}{2}\Omega\right)$ identically corresponds one and only one term of $\tilde{G}_{-\mu}^{(n)}\left(\omega+\frac{\mu}{2}\Omega\right)$. This allows to conclude that $\tilde{G}_{\mu}^{(n)}\left(\omega-\frac{\mu}{2}\Omega\right)=\tilde{G}_{-\mu}^{(n)}\left(\omega+\frac{\mu}{2}\Omega\right)$ is valid  $\forall n$. By virtue of the series expansion in Eq.~(\ref{app_eq:series}), the above property is valid also for the complete $\tilde{G}_\mu(\omega)$:
\be
\tilde{ G}_{\mu}\left(\omega-\frac{\mu}{2}\Omega\right)=\tilde{ G}_{-\mu}\left(\omega+\frac{\mu}{2}\Omega\right)\,.\nonumber
\ee

\section{Non--Markovianity criterion}
\label{appendix4}
Memory effects and non--Markovian dynamics  are the subject of many studies, and several notions of non--Markovianity have been introduced recently (see the review in Ref.~\onlinecite{breuerrmp} and references therein). Different estimators have been investigated {to witness} and to quantify non--Markovianity, and these not always are equivalent~\cite{breuerrmp}. Moreover, many criteria, such as the ones related to the trace distance, are not well-suited in the study of asymptotic states, i.e. looking at properties in the long time limit (like in our case of interest). There, different estimators should be used, as discussed in Refs. ~\cite{rivas10, groeblacher, illuminatiprl, illuminatipra}. In particular, it has been shown that non--Markovianity can be assessed through the violation of the divisibility condition: 
the evolution is  Markovian if and only if it is described by a divisible, completely positive map. Importantly, this criterion is valid for both finite and asymptotic times.\\
In Refs.~\cite{groeblacher, illuminatiprl, illuminatipra} this criterion has been applied to the case of a QHO coupled to a thermal bath at temperature $T$, exploiting the exact solution  derived in Ref.~\cite{hu92}, and the divisibility estimator has been linked directly to the form of the bath spectral density $\mathcal{J}(\omega)$. We now recall the definition of this estimator, details on the derivation can be found in Refs.~\cite{groeblacher, illuminatipra}.
The punctual non--Markovianity measure is given by
\be
\label{app_nt}
{\cal N}_p(t) = \frac{1}{2}\left[1-\frac{\Delta(t)}{\sqrt{\Delta^2(t) + \Gamma^2(t) + \Pi ^2(t)}}\right]
\ee
where $\Gamma(t)$, $\Delta(t)$, and $\Pi(t)$ are the damping, direct and anomalous diffusion coefficients, respectively, and are completely determined by the form of the bath spectral density ${\cal J}(\omega)$.
 Notice that we have indicated the damping coefficient with $\Gamma(t)$, instead of $\gamma(t)$ used in the original paper \cite{illuminatipra} to avoid confusion with the quantity introduced in the main text. Equation~\eqref{app_nt} is valid at any time, and in particular also in the asymptotic $t\to\infty$ regime. It is worth to note that this quantity is bounded $0\leq {\cal N}_p(t) \leq 1$.
If ${\cal N}_p(t)=0$, one has a Markovian dynamics, hence described by a separable dynamical map. On the other hand, if ${\cal N}_p(t)>0$ the divisibility criterion is violated and the dynamics is non--Markovian. In general, the above three time-dependent coefficients are given by cumbersome integral expressions.
However, at weak coupling the above coefficients can be evaluated in closed form, and in the asymptotic regime they read~\cite{groeblacher, illuminatipra}
\be
\Gamma(\infty)= \frac{{\cal J}(\omega_0)}{2m\omega_0}\quad,\quad\Delta(\infty)= \Gamma(\infty)\coth\left(\frac{\omega_0}{2T}\right)
\ee
and
\be
\Pi(\infty)= - {\cal P} \int_0^{+\infty}\mathrm{d}\omega \frac{{\cal J}(\omega)}{\pi m} \coth\left(\frac{\omega}{2T}\right)\frac{1}{\omega^2-\omega_0^2}\label{app_eq:Piill}
\ee
where the integral is taken as a principal value.
We underline that, since $\Delta(\infty)$ is positive definite, in the asymptotic regime ${\cal N}_p(\infty)$ is bounded between $0$ (Markovian regime) and the maximal non--Markovianity $1/2$.
Hereafter, we analyze the behavior of ${\cal N}_p(\infty)$ for the case of baths with a spectral density $\propto\omega^s$ ($s>0$) or with a structured Lorentzian spectral density, which are the examples discussed in the main text. We will show in particular that only the strictly Ohmic spectral density at high temperature (classical regime) leads to ${\cal N}_p(\infty)=0$ and thus to a Markovian dynamics.\\
\begin{figure}
\includegraphics[width=8cm,keepaspectratio]{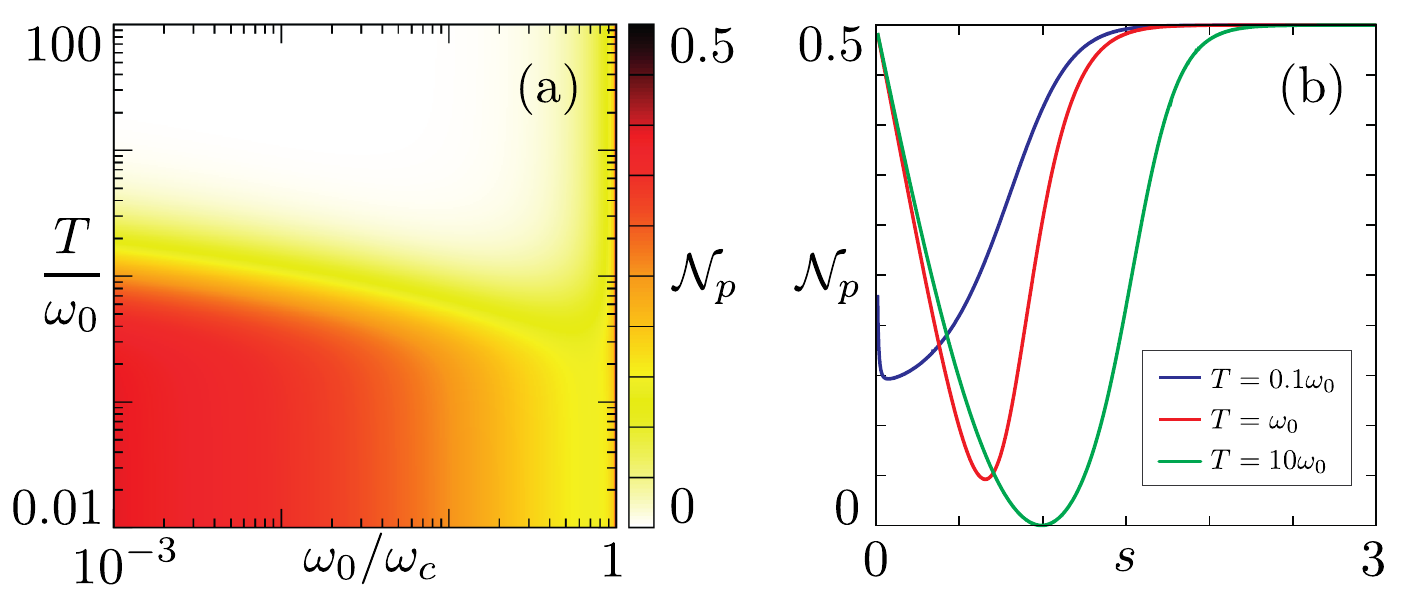}
\caption{Punctual non--Markovianity measure in the asymptotic regime ${\mathcal{N}}_p$ for a spectral density $\propto\omega^s$ with a hard cutoff $\omega_c$. (a) Density plot of ${\mathcal{N}}_p={\mathcal{N}}_p(\infty)$ as a function of the inverse cutoff $\omega_0/\omega_c$ and $T/\omega_0$ for the Ohmic $s=1$ case. (b) Plot of ${\mathcal{N}}_p$ as a function of $s$ for three different temperatures (see legend) and $\omega_c=10^4\omega_0$.}\label{app_fig:ill1}	
\end{figure}

We begin considering the spectral density 
\begin{equation}
\mathcal{J}(\omega)=m\gamma\omega\left|\frac{\omega}{\bar{\omega}}\right|^{s-1}\theta(\omega_c-|\omega|)\,,   
\end{equation}
where $\theta(x)$ is the Heaviside theta function, which models a sub--Ohmic ($0<s<1$), Ohmic ($s=1$) or super--Ohmic ($s>1$) bath with a hard cutoff $\omega_c$. Figure~\ref{app_fig:ill1}(a) shows ${\cal N}_p={\cal N}_p(\infty)$ for the Ohmic $s=1$ case, as a function of the (inverse) cutoff and temperature. It is clear that {\em only} for $\omega_c\gg\omega_0$ and $T\gg\omega_0$ one has ${\cal N}_p\to 0$. Indeed, setting $T\gg\omega_0$ and evaluating the principal value in Eq.~(\ref{app_eq:Piill}) one finds $\Pi(\infty)\equiv 0$, while in the same regime $\Gamma(\infty)/\Delta(\infty)=\frac{\omega_0}{2T}\to 0$, which proves that ${\cal N}_p\propto\left(\frac{\omega}{T}\right)^2\to 0$. Away from this regime deviations from the Markovianity occur, particularly in the $T<\omega_0$ (quantum) regime. It is also worth investigating the sub-- and super--Ohmic cases for large cutoff ($\omega_c\gg\omega_0$). The results are summarized in Fig.~\ref{app_fig:ill1}(b) where ${\cal N}_p$ is shown as a function of $s$ for three different temperatures. In all cases except the high--temperature Ohmic one, the dynamics is non--Markovian.\\
\begin{figure}
\includegraphics[width=8cm,keepaspectratio]{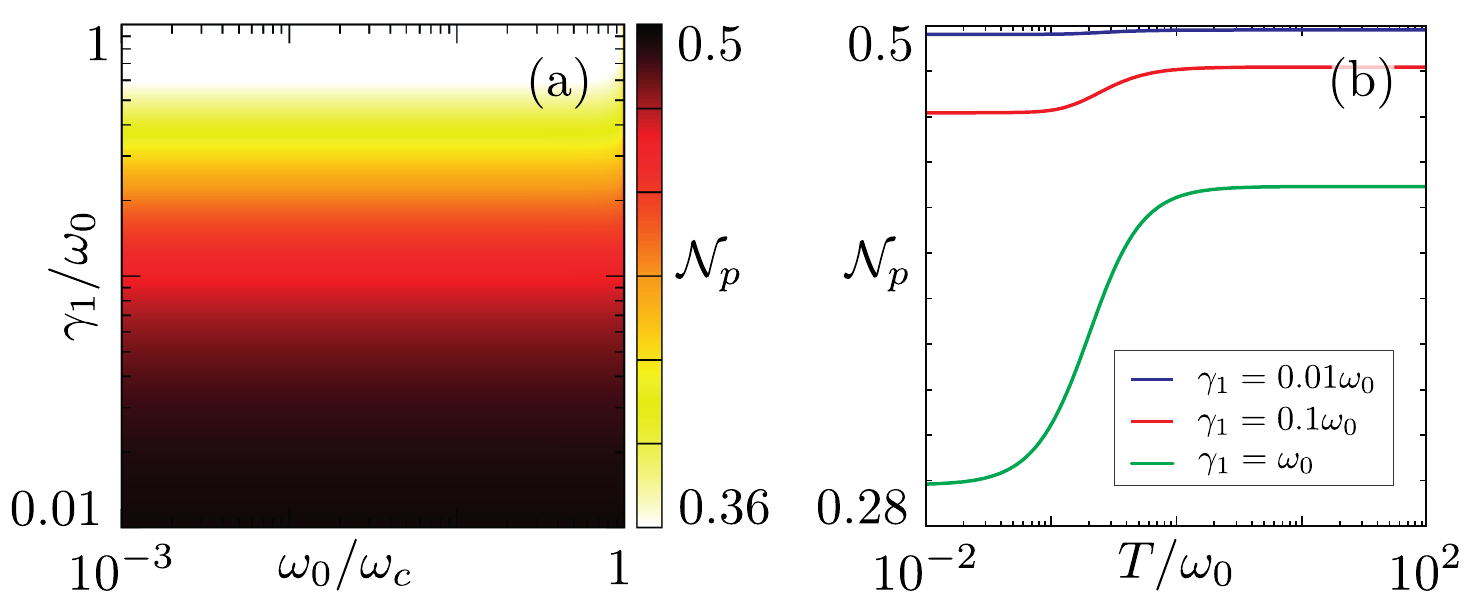}
\caption{Punctual non--Markovianity measure in the asymptotic regime ${\mathcal{N}}_p={\mathcal{N}}_p(\infty)$ for a Lorentzian spectral density with a hard cutoff. (a) Density plot of ${\mathcal{N}}_p$ as a function of the inverse cutoff $\omega_0/\omega_c$ and $\gamma_1$ for $\omega_1=\omega_0/2$ and $T=0.2\omega_0$. (b) Plot of ${\mathcal{N}}_p$ as a function of $T/\omega_0$ for three different values of $\gamma_1$ (see legend) and $\omega_c=10^4\omega_0$.}\label{app_fig:ill2}	
\end{figure}

We now turn to the case of a Lorentzian spectral density with a hard cut--off 
\begin{equation}
{\cal J}(\omega)=\frac{m d_1\gamma_1\omega\theta(\omega_c-|\omega|)}{\left(\omega^2-\omega_1^2\right)^2+\gamma_1^2\omega^2}\,,   
\end{equation}
discussed in Sec.\ref{sec:NMengine} in the large cut--off limit.
Figure~\ref{app_fig:ill2}(a) shows the non--Markovianity measure as a function of the cutoff and the damping parameter $\gamma_1$ for typical values of $\omega_1$ and $T$. In this case we found for $\mathcal{N}_p$ a minimum value $\approx 0.36$ for a broad Lorentzian peak ($\gamma_1\approx\omega_0$) which increases to the maximum $\mathcal{N}_p=1/2$ when the peak is very sharp ($\gamma_1\ll\omega_0$), signaling a distinctly non--Markovian dynamics. Also, it is worth to notice that ${\cal N}_p$ is essentially insensitive to the cutoff when $\omega_c>\omega_1$. Figure~\ref{app_fig:ill2}(b) confirms that the non--Markovian dynamics is stable against thermal effects.

\section{Weak coupling}
\label{appendix5}
In this Section we derive closed expressions for the average power and heat currents in the weak coupling regime with respect to ${\cal J}_1(\omega)$. The starting point are the general forms previously obtained for the power in Eq.~(\ref{Paverage}) and for the heat current 
in Eq.~(\ref{app_sjaaverage}). We first need the perturbative expansion of the Floquet coefficients in Eq.~\eqref{app_sfloquet1}, which, recalling Eqs.~(\ref{app_s1kappan1}) and~(\ref{app_s1kappan2}), can be written up to linear order in ${\cal J}_1$ as
\beq
\label{app_ssfloquet}
&&\tilde{G}_0(\omega)= \chi_0(\omega)\left[1+\frac{i}{4}[\omega_+\tilde\gamma_1(\omega_+)+\omega_-\tilde\gamma_1(\omega_-)]
\chi_0(\omega)\right], \nonumber \\
&&\tilde{G}_{\pm2}(\omega)=\frac{i}{4}\chi_0(\omega \pm 2 \Omega)\omega_{\pm} \tilde\gamma_1(\omega_{\pm})
\chi_0(\omega), \nonumber \\
&& \tilde{G}_{|m|>2}={\cal O}({\cal J}_1^2)\,.
\eeq
Above we have introduced the bare susceptivity
\be
{\chi}_0(\omega)=-\frac{1}{\omega^2-\omega_0^2+i\omega\tilde\gamma_2(\omega)}~,
\ee
with effective damping $\tilde{\gamma}_2(\omega)$. Notice that this quantity fulfill the relation
\be
\label{app_grelation}
\!\!\!\!\!\!\Im [\chi_0(\omega)]\!\!=\!\!\omega|\chi_0(\omega)|^2\Re [\tilde\gamma_2(\omega)]\!\!=\!\!\frac{1}{m}
{\cal J}_2(\omega)|\chi_0(\omega)|^2~.
\ee
Considering the average power in Eq.~(\ref{Paverage}), we should evaluate Eqs.~\eqref{app_p1afinal},~\eqref{app_P1b1final} and ~\eqref{app_P1b2final} up to linear order in ${\cal J}_1$. The first contribution is already linear in ${\cal J}_1$, hence, by using the zero-th term $\tilde{G}_0(\omega)=\chi_0(\omega)$ in Eq.~\eqref{app_ssfloquet} one has
\begin{equation*}
P^{(a)}=-\frac{\Omega}{4\pi m}\int_{-\infty}^{+\infty}\!\!\mathrm{d}\omega{\cal J}_1(\omega)\coth(\frac{\omega}{2T_1})  \Im\left[\chi_{0}(\omega-\Omega)\right]~.
\end{equation*}
Equation~\eqref{app_P1b1final} does not contribute, since it is at least of second order in ${\cal J}_1$. Then, using Eq.~\eqref{app_grelation} we can write the first order contribution of Eq.~\eqref{app_P1b2final} as
\begin{eqnarray}
P^{(b,2)}&=&\frac{\Omega}{8\pi m}\int_{-\infty}^{+\infty}\!\!\!\mathrm{d}\omega\coth(\frac{\omega}{2T_{2}})\nonumber\\
&\times&\Im[\chi_{0}(\omega)]\Big[{\cal J}_{1}(\omega+\Omega)-{\cal J}_{1}(\omega-\Omega)\Big].\label{app_P1b2weak}
\end{eqnarray}
Combining the above expressions and recalling that $\coth(x/2)=1+ 2 n_B(x)$, with $n_B(x)=(e^x-1)^{-1}$, we arrive at
\begin{widetext}
\be
P=-\Omega\int_{0}^{+\infty}\!\!\frac{\mathrm{d}\omega}{2\pi m}\Im\chi_0(\omega)\Big[ {\cal J}_1 (\omega_+) n_B \left(\frac{\omega_+}{T_1} \right) -
  {\cal J}_1 (\omega_-) n_B \left(\frac{\omega_-}{T_1} \right)+ [{\cal J}_1 (\omega_-)-{\cal J}_1 (\omega_+)] n_B \left(
  \frac{\omega}{T_2} \right)\Big]. \label{app_eq:Ppert}
  \ee
\end{widetext}
We now focus on the average current $J_2$ as reported in Eq.~\eqref{app_sjaaverage}, with Eqs.~\eqref{app_J2a},~\eqref{app_J2bbis}. To this end, we can use the following relations (valid up to linear order in ${\cal J}_1$):
\begin{widetext}
\be
|\tilde{G}_0(\omega)|^2=|\chi_0(\omega)|^2\left[1-\frac{1}{2}\sum_{p=\pm}\omega_p \Im\Big(\chi_0(\omega)\tilde{\gamma}_1(\omega_p)\Big)\right]\,,
\ee
\be
\Im\tilde{G}_0(\omega)=\Im\chi_0(\omega)+\frac{1}{4}\sum_{p=\pm}
\omega_p\Big[\Re\Big(\chi_0(\omega)\Big)\Re\Big(\chi_0(\omega)\tilde{\gamma}_1(\omega_p)\Big)-\Im\Big(\chi_0(\omega)\Big)\Im\Big(\chi_0(\omega)\tilde{\gamma}_1(\omega_p)\Big)\Big]\,.
\ee
\end{widetext}
Comparing the two above equations and recalling Eq.~\eqref{app_grelation} we can write
\be
\!\!\frac{{\cal J}_2(\omega)}{m}|\tilde{G}_0(\omega)|^2\!\!=\!\!\Im\tilde{G}_0(\omega)\!\!-\!\!\frac{1}{4m}|\chi_0(\omega)|^2
\sum_{p=\pm}{\cal J}_1(\omega_{p}).
\ee
Plugging these expressions into Eq.~\eqref{app_J2a},~\eqref{app_J2bbis} and using the explicit form of $\tilde{G}_{\pm2}(\omega)$ in Eq.~\eqref{app_ssfloquet} (linear in ${\cal J}_1$) we arrive 
at the compact form
\begin{widetext}
\beq
&&J_2=-\int_{-\infty}^{+\infty}\!\!\frac{\mathrm{d}\omega}{4\pi m}\omega{\cal J}_{1}(\omega+\Omega)\Im\chi_0(\omega)\left[\coth(\frac{\omega+\Omega}{2T_1})-\coth(\frac{\omega}{2T_{2}})\right],\label{app_eq:J2pert}\\
&&J_1=\int_{-\infty}^{+\infty}\!\!\frac{\mathrm{d}\omega}{4\pi m}(\omega+\Omega){\cal J}_{1}(\omega+\Omega)\Im\chi_0(\omega)\left[\coth(\frac{\omega+\Omega}{2T_1})-\coth(\frac{\omega}{2T_{2}})\right],\label{app_eq:J1pert}
\eeq
\end{widetext}
where, for the sake of completeness, we have quoted also the expression for $J_1=-(P+J_2)$.

\section{No heat engine for Ohmic and sub-Ohmic spectral function}
\label{appendix6}

Here we show that when $\mathcal{J}_1(\omega)=m\gamma_1\omega\left|\frac{\omega}{\bar{\omega}}\right|^{s-1}$ and $0<s\leq 1$ no working heat engine can be achieved. Note that in this Section we assume that the cut-off $\omega_c$ is the largest energy scale ($\omega_c \gg \omega_0, \Omega$). To somewhat ease the proof, we focus on the case of a Ohmic bath for the static contact 2, with $\tilde{\gamma}_2(\omega)=\gamma_2\ll\omega_0$. In this regime the imaginary part of the bare susceptibility is well described by
\be
\mathrm{Im}\{\chi_0(\omega)\}\approx\frac{\pi}{2\omega_0}\left[\delta(\omega-\omega_0)-\delta(\omega+\omega_0)\right]\, .\label{app_eq:deltas}
\ee
Plugging Eq.~(\ref{app_eq:deltas}) into Eqs.~(\ref{app_eq:Ppert}),~(\ref{app_eq:J1pert}) and~(\ref{app_eq:J2pert}) the average power and heat currents reduce to
\begin{widetext}
\beq
&&P=-\frac{\Omega}{4m\omega_0}\sum_{p=\pm1}p\mathcal{J}_1(\omega_0+p\Omega)\left[n_B\left(\frac{\omega_0+p\Omega}{T_1}\right)-n_B\left(\frac{\omega_0}{T_2}\right)\right]=\sum_{p=\pm1}P^{(p)}\,,\label{app_eq:ppd}\\
&&J_1=\frac{1}{4m}\sum_{p=\pm1}\left(\frac{\omega_0+p\Omega}{\omega_0}\right)\mathcal{J}_1(\omega_0+p\Omega)\left[n_B\left(\frac{\omega_0+p\Omega}{T_1}\right)-n_B\left(\frac{\omega_0}{T_2}\right)\right]=\sum_{p=\pm1}J_1^{(p)}\,,\label{app_eq:j1pd}\\
&&J_2=-\frac{1}{4m}\sum_{p=\pm1}\mathcal{J}_1(\omega_0+p\Omega)\left[n_B\left(\frac{\omega_0+p\Omega}{T_1}\right)-n_B\left(\frac{\omega_0}{T_2}\right)\right]=\sum_{p=\pm1}J_2^{(p)}\,.\label{app_eq:j2pd}
\eeq
\end{widetext}

 We start observing that given Eqs.~(\ref{app_eq:deltas}) and~(\ref{finalP}) the condition in Eq.~(\ref{eq:thecondition}) becomes $f(\omega_0,\Omega)>0$. The best--case scenario, for this monotonic spectral density, occurs when $T_2\to 0$, as also discussed in the main text, since it minimizes for given $s,\Omega,T_1$ the positive contribution $\propto[\mathcal{J}_1(\omega_0+\Omega)-\mathcal{J}_1(\omega_0-\Omega)]n_B\left(\frac{\omega_0}{T_2}\right)$ to the power $P$ and thus maximizes the power output. The condition $f(\omega_0,\Omega)>0$ is then equivalent to
\begin{equation}
\sum_{p=\pm1}p(\omega_0+p\Omega)|\omega_0+p\Omega|^{s-1}n_B\left(\frac{\omega_0+p\Omega}{T_1}\right)>0\, .
\end{equation}
Observing that $(\omega_0-\Omega)n_B\left(\frac{\omega_0-\Omega}{T_1}\right)>0$ always, with some rearrangements the above equation becomes 
\be
\label{app_eq:noh1}
f_s(\Omega) > g_{T_1}(\Omega)
\ee
where 
\beq
f_s(\Omega) &=& \Big|\frac{\omega_0+\Omega}{\omega_0-\Omega}\Big|^s \label{app_eq:fs} \\
g_{T_1}(\Omega)&=& \Big|\frac{n_B ((\omega_0-\Omega)/T_1)}{n_B((\omega_0+\Omega)/T_1)}\Big|\label{app_eq:gT}\, .
\eeq
We firstly observe that both $f_s(\Omega)$ and $g_{T_1}(\omega)$ are continuous functions of $\Omega$ except at $\Omega=\omega_0$ where they diverge as 
\begin{equation}
f_s(\Omega)\approx\frac{(2\omega_0)^s}{|\omega_0-\Omega|^s}\ ;\  g_{T_1}(\Omega)\approx\frac{T_1\left(e^{2\omega_0/T_1}-1\right)}{|\omega_0-\Omega|}
\end{equation}
when $\Omega\to\omega_0$. At least for $\Omega\approx\omega_0$ and $0<s\leq 1$ it is then clear that $f_s(\Omega)\leq g_{T_1}(\Omega)$ and thus no working engine can be obtained there. To prove that this is the case for any $\Omega$  we now inspect the general properties of $f_s(\Omega)$ and $g_{T_1}(\Omega)$, and their derivatives, to show that for $0<s\leq 1$ Eq.~\eqref{app_eq:noh1} cannot be verified. Observe that
\begin{equation}
\label{app_dfs}
\frac{d f_s(\Omega)}{d\Omega}=s\left(\frac{2\omega_0}{\omega_0^2-\Omega^2}\right)f_s(\Omega)\quad\mathrm{for}\ {\Omega\neq\omega_0}\ ,
\end{equation}
and
\be
\frac{d g_{T_1}(\Omega)}{d\Omega}=\phi_{T_1}(\Omega)g_{T_1}(\Omega)~,
\ee
where
\beq
\phi_{T_1}(\Omega) = \begin{cases}
\frac{1 +\delta_{T_1}(\Omega)}{T_1} & {\rm if}\ 0 < \Omega < \omega_0 \\
\frac{\delta_{T_1}(\Omega)}{T_1} & {\rm if}\ \Omega > \omega_0
\end{cases}
\eeq
with $\delta_{T_1}(\Omega)=\sinh(\omega_0/{T_1})/[\cosh(\omega_0/{T_1})-\cosh(\Omega/{T_1})]$. It can be checked (not shown here) that $\phi_{T_1}(\Omega)$ is a monotonically decreasing function of $T_1$ and that
\begin{equation}
\lim_{T_1\to\infty}\phi_{T_1}(\Omega)=\frac{2\omega_0}{\omega_0^2-\Omega^2}=\frac{1}{s f_s(\Omega)}\frac{d f_s(\Omega)}{d\Omega}~,
\end{equation}
where in the second passage we have used Eq.~\eqref{app_dfs}. Thus we arrive at the following inequality
\begin{equation}
\frac{d g_{T_1}(\Omega)}{d \Omega}>\frac{1}{s}\frac{g_{T_1}(\Omega)}{f_s(\Omega)}\frac{d f_s(\Omega)}{d\Omega}\,.\label{app_eq:der2}
\end{equation}
Integrating Eq.~\eqref{app_eq:der2} from $0$ to $\Omega$ one obtains $f_s(\Omega)<g_{T_1}^s(\Omega)$ or equivalently
\begin{equation}
\frac{f_s(\Omega)}{g_{T_1}(\omega)}<g_{T_1}^{s-1}(\Omega).
\end{equation}
Since $g_{T_1}(\Omega)>1$ for $\Omega>0$, from the last inequality it follows that if $0<s\leq 1$ one has $f_s(\Omega)<g_{T_1}(\Omega)$. This is in contrast with Eq.~(\ref{app_eq:noh1}) and this shows that $f(\omega_0,\Omega)<0$ and then no engine can be achieved for $0<s\leq 1$.


\begin{thebibliography}{177}
\bibitem{pekola15} J. P. Pekola, Towards quantum thermodynamics in electronic circuits, Nature Phys. 11, 118 (2015).

\bibitem{vinjanampathy2015}
S. Vinjanampathy and J. Anders, Quantum Thermodynamics, Contemporary Physics \textbf{57},  1 (2016).
\bibitem{benenti17} G Benenti, G Casati, K Saito, and R. S. Whitney, Fundamental aspects of steady-state conversion of heat to work at the nanoscale, Phys. Rep. \textbf{694}, 1 (2017).
\bibitem{curzon} 
	F. Curzon and B. Alhborn, Efficiency of a Carnot engine at maximum power output, Am. J. Phys. {\bf 43}, 22 (1975).
\bibitem{campaiolibook} F. Campaioli, F. A. Pollock, and S. Vinjanampathy, \emph{Thermodynamics in the Quantum Regime}, edited by F. Binder, L. A. Correa, C. Gogolin, J. Anders, and G. Adesso, (Springer, Berlin, 2018).
\bibitem{killoran}
N. Killoran, S. F. Huelga, and M. Plenio, Enhancing light-harvesting power with coherent vibrational interactions: a quantum heat engine picture, J. Chem. Phys. {\bf 143}, 155102 (2015).
\bibitem{ciliberto}
S. Ciliberto, Experiments in Stochastic Thermodynamics: Short History and Perspectives, Phys. Rev. X {\bf 7}, 021051 (2017).
\bibitem{bouton}
Q. Bouton, J. Nettersheim, S. Burgardt, D. Adam, E. Lutz, and A. Widera, A quantum heat engine driven by atomic collisions, Nat. Commun. {\bf 12}, 2063 (2021).
\bibitem{thierschmann15} H. Thierschmann, R. S\'anchez, B. Sothmann, F. Arnold, C. Heyn, W. Hansen, H. Buhmann, and L. W. Molenkamp, Three-terminal energy harvester with coupled quantum dots, Nature Nanotech. \textbf{10}, 854 (2015). 
\bibitem{martinez16} I. A. Martinez, \'E. Rold\'an, L. Dinis, D. Petrov, J. M. R. Parrondo, and R. A. Rica, Brownian Carnot Engine, Nat. Phys. {\bf 12}, 67 (2016).
\bibitem{blasi21} G. Blasi, F. Taddei, L. Arrachea, M. Carrega, and A. Braggio, Nonlocal thermoelectric engines in hybrid topological Josephson junctions, Phys. Rev. B {\bf 103}, 235434 (2021).
\bibitem{xu22} M. Xu, J. T. Stockburger, G. Kurizki, and J. Ankerhold, Minimal quantum thermal machine in a bandgap environment: non-Markovian features and anti-Zeno advantage, New J. Phys. {\bf 24}, 035003 (2022). 
%
\bibitem{esposito09}
M. Esposito, U. Harbola, and S. Mukamel, Nonequilibrium fluctuations, fluctuation theorems, and counting statistics in quantum systems, { Rev. Mod. Phys.}  \textbf{81}, 1665 (2009).
  
\bibitem{campisi11}
M. Campisi, P. H{\"a}nggi, and P. Talkner, {\it Colloquium}: Quantum fluctuation relations: Foundations and applications, { Rev. Mod. Phys.} \textbf{83}, 771 (2011).
  
\bibitem{kosloff13}
R. Kosloff, Quantum Thermodynamics: A Dynamical Viewpoint, Entropy \textbf{15}, 2100 (2013).

\bibitem{goold2016}
J. Goold, M. Huber, A. Riera, L. del Rio, and P. Skrzypczyk, The role of quantum information in thermodynamics—a topical review, J. Phys. A Math. Theor. \textbf{49}, 1430001 (2016).

\bibitem{carrega19} M. Carrega, M. Sassetti, and U. Weiss, Optimal work-to-work conversion of a nonlinear quantum Brownian duet, Phys. Rev. A \textbf{99}, 062111 (2019).
\bibitem{miller19} H. J. D. Miller, M. Scandi, J. Anders, and M. Perarnau-Llobet, Work Fluctuations in Slow Processes: Quantum Signatures and Optimal Control, Phys. Rev. Lett. {\bf 123}, 230603 (2019).

\bibitem{khandelwal} S. Khandelwal, N. Palazzo, N. Brunner, and G. Haack, Critical heat current for operating an entanglement engine,  New J. Phys. {\bf 22}, 073039 (2020).
\bibitem{vischi19} F. Vischi, M. Carrega, P. Virtanen, E. Strambini, A. Braggio, and F. Giazotto, Thermodynamic cycles in Josephson junctions, Sci. Rep. {\bf 9}, 3238 (2019).
\bibitem{bhandari21} B. Bhandari, P. A. Erdman, R. Fazio, E. Paladino, and F. Taddei, Thermal rectification through a nonlinear quantum resonator, Phys. Rev. B {\bf 103}, 155434 (2021).
\bibitem{pancotti20} N. Pancotti, M. Scandi, M. T. Mitchison, and M. Perarnau-LLobet, Speed-Ups to Isothermality: Enhanced Quantum Thermal Machines through Control of the System-Bath Coupling, Phys. Rev. X {\bf 10}, 031015 (2020).

\bibitem{jurgen} M. Wiedmann, J. T. Stockburger, and J. Ankerhold, Non-Markovian dynamics of a quantum heat engine: out-of- equilibrium operation and thermal coupling control, New J. Phys. \textbf{22}, 033007 (2020).

\bibitem{esposito19} K. Ptaszynski and M. Esposito, Entropy Production in Open Systems: The Predominant Role of Intraenvironment Correlations, Phys. Rev. Lett. {\bf 123}, 200603 (2019).

\bibitem{liu22} J. Liu, K. A. Jung, and D. Segal, Periodically Driven Quantum Thermal Machines from Warming up to Limit Cycle, Phys. Rev. Lett. {\bf 127}, 200602 (2022).
\bibitem{brandner166} K. Brandner and U. Seifert, Periodic thermodynamics of open quantum systems, Phys. Rev. E {\bf 93}, 062134 (2016).
\bibitem{son22} J. Son, P. Talkner, and J. Thingna, Monitoring Quantum Otto Engines, Phys. Rev. X Quantum {\bf 2}, 040328 (2021).
%
\bibitem{breuerbook} H.-P. Breuer and F. Petruccione, \textit{The theory of open quantum systems} (Oxford University Press, 2002).
\bibitem{weiss} U. Weiss, \textit{Quantum Dissipative Systems - 5th Edition} (World Scientific, Singapore, 2021).
\bibitem{aurel18} E. Aurell, Characteristic functions of quantum heat with baths at different temperatures, Phys. Rev. E {\bf 97}, 062117 (2018).
%
\bibitem{strasberg16} P. Strasberg, G. Schaller, N. Lambert, and T. Brandes, Nonequilibrium thermodynamics in the strong coupling and non-Markovian regime based on a reaction coordinate mapping, New. J. Phys. {\bf 18}, 073007 (2016).
\bibitem{restrepo18} S. Restrepo, J. Cerrillo, P. Strasberg, and G. Schaller, 
From quantum heat engines to laser cooling: Floquet theory beyond the Born–Markov approximation, New J. Phys {\bf 20}, 053063 (2018).

\bibitem{qcbook} G. Benenti, G. Casati, D. Rossini, and G. Strini, \textit{Principles of quantum computation and information (A comprehensive textbook)} (World Scientific, Singapore, 2019).

\bibitem{wang} W. Wang, J. Han, B. Yadin, Y. Ma, J. Ma, W. Cai, Y. Xu, L. Hu, H-. Wang, Y. P. Song, M. Gu, and L. Sun, Witnessing Quantum Resource Conversion within Deterministic Quantum Computation Using One Pure Superconducting Qubit, Phys. Rev. Lett. {\bf 123}, 220501 (2019).
\bibitem{elouard} C. Elouard, G. Thomas, O. Maillet, J. P. Pekola, and A. N. Jordan, Quantifying the quantum heat contribution from a driven superconducting circuit, Phys. Rev. E {\bf 102}, 030102(R) (2020).
\bibitem{rossini} D. Rossini, G. M. Andolina, D. Rosa, M. Carrega, and M. Polini, Quantum Advantage in the Charging Process of Sachdev-Ye-Kitaev Batteries, Phys. Rev. Lett. {\bf 125}, 236402 (2020).
\bibitem{gyhm} J.-Y. Gyhm, D. \v{S}afranek, and D. Rosa, Quantum Charging Advantage Cannot Be Extensive without Global Operations, Phys. Rev. Lett. {\bf 128}, 140501 (2022).
\bibitem{watanabe0} G. Watanabe, B. Prasanna Venkatesh, P. Talkner, and A. Del Campo, Quantum Performance of Thermal Machines over Many Cycles, Phys. Rev. Lett. {\bf 118}, 050601 (2017).

\bibitem{watanabe} G. Watanabe, B. Prasanna Venkatesh, P. Talkner, M.-J. Wang, and A. Del Campo, Quantum Statistical Enhancement of the Collective Performance of Multiple Bosonic Engines, Phys. Rev. Lett. {\bf 124}, 210603 (2020).
\bibitem{hammam} K. Hammam, H. Leitch, Y. Hassouni, and G. De Chiara, Exploiting coherence for quantum thermodynamic advantage, ArXiv:2202.07515 (2022).
%
\bibitem{talkner} P. Talkner and P. H\"anggi, {\it Colloquium}: Statistical mechanics and thermodynamics at strong coupling: Quantum and classica, Rev. Mod. Phys. {\bf 92}, 41002 (2020).
\bibitem{moskalets} M. F. Ludovico, L. Arrachea, M. Moskalets, and D. Sanchez, Periodic Energy Transport and Entropy Production in Quantum Electronics, Entropy {\bf 18}, 419 (2016).
\bibitem{paternostro} G. T. Landi and M. Paternostro, Irreversible entropy production: From classical to quantum, Rev. Mod. Phys. {\bf 93}, 035008 (2021).


\bibitem{devega} I. de Vega and D. Alonso, Dynamics of non-Markovian open quantum systems, Rev. Mod. Phys. \textbf{89}, 015001 (2017).
\
\bibitem{landi_prxquantum} G. T. Landi, M. Paternostro, and A. Belenchia, Informational Steady States and Conditional Entropy Production in Continuously Monitored Systems, PRX Quantum {\bf 3}, 010303 (2022).
\bibitem{hewgill} A. Hewgill, G. De Chiara, and A. Imparato, Quantum thermodynamically consistent local master equations, Phys. Rev. Research {\bf 3}, 013165 (2021).
\bibitem{carrega_prxquantum} M. Carrega, L. M. Cangemi, G. De Filippis, V. Cataudella, G. Benenti, and M. Sassetti, Engineering Dynamical Couplings for Quantum Thermodynamic Tasks, PRX quantum {\bf 3}, 010323 (2022). 

\bibitem{ivander22} F. Ivander, N. Anto-Sztrikacs, and D. Segal, Strong system-bath coupling reshapes characteristics of quantum thermal machines, ArXiv:2111.05302 (2021).

\bibitem{shirai} Y. Shirai, K. Ashimoto, R. Tezuka, K. Uchiyama, and N. Atano, Non-Markovian effect on quantum Otto engine: Role of system-reservoir interaction, Phys. Rev. Research {\bf 3}, 023078 (2021).
\bibitem{breuerrmp} H. P. Breuer, E.-M. Laine, J. Piilo, and B. Vacchini, {\it Colloquium}: Non-Markovian dynamics in open quantum systems, Rev. Mod. Phys. {\bf 88}, 021002 (2016).
\bibitem{nori12} W.-M. Zhang, P.J. Lo, H.-N. Xiong, M. W.-J. Tu, and F. Nori, General Non-Markovian Dynamics of Open Quantum Systems, Phys. Rev. Lett. {\bf 109}, 170402 (2012).
\bibitem{newpt} K. Ptaszynski, Non-Markovian thermal operations boosting the performance of quantum heat engines, Phys. Rev. E {\bf 106}, 014114 (2022).

%
\bibitem{leitch} H. Leitch, N. Piccione, B. Bellomo, and G. De Chiara, Driven quantum harmonic oscillators: A working medium for thermal machines, AVS Quantum Sci. {\bf 4}, 012001 (2022).
%
\bibitem{hofer} P. P. Hofer, M. Perarnau-LLobet, L. D. M. Miranda, G. Haack, R. Silva, J. Bohr Brask, and N. Brunner, Markovian master equations for quantum thermal machines: local vs global approach, New J. Phys. {\bf 19}, 123037 (2017).
\bibitem{haake} F. Haake and R. Reibold, Strong damping and low-temperature anomalies for the harmonic oscillator, Phys. Rev. A {\bf 32}, 2462 (1985).

\bibitem{hu92} B. L. Hu, J. P. Paz, and Y. Zhang, Quantum Brownian motion in a general environment: Exact master equation with nonlocal dissipation and colored noise, Phys. Rev. D {\bf 45}, 2843 (1992).

\bibitem{CL83} A. O. Caldeira and A. J. Leggett, Quantum tunnelling in a dissipative system, Ann. Phys. \textbf{149}, 374 (1983).

\bibitem{cangemi} L. M. Cangemi, M. Carrega, A. De Candia, V. Cataudella, G. De Filippis, M. Sassetti, and G. Benenti, Optimal energy conversion through antiadiabatic driving breaking time-reversal symmetry, Phys. Rev. Research \textbf{3}, 013237 (2021). 
\bibitem{flindt21} P. Portugal, C. Flindt, and N. Lo Gullo, Heat transport in a two-level system driven by a time-dependent temperature, Phys. Rev. B {\bf 104}, 205420 (2021).
\bibitem{paz1} N. Freitas and J. P. Paz, Fundamental limits for cooling of linear quantum refrigerators, Phys. Rev. E \textbf{95}, 012146 (2017).
\bibitem{paz2} N. Freitas and J. P. Paz, Cooling a quantum oscillator: A useful analogy to understand laser cooling as a thermodynamical process, Phys. Rev. A \textbf{97}, 032104 (2018).
\bibitem{arrachea12a} L. Arrachea, E. Mucciolo, C. Chamon, and R. B. Capaz, Microscopic model of a phononic refrigerator, Phys. Rev. B {\bf 86}, 125424 (2012).

\bibitem{grifoni95} M. Grifoni, M. Sassetti, P. Hanggi, and U. Weiss, Cooperative effects in the nonlinearly driven spin-boson system, Phys. Rev. E \textbf{52}, 3596 (1995).
\bibitem{rivas10} A. Rivas, S. F. Huelga, and and M. B. Plenio, Entanglement and Non-Markovianity of Quantum Evolutions, Phys. Rev. Lett. {\bf 105}, 050403 (2010).
\bibitem{groeblacher} S. Gr\"oblacher, A. Trubarov, N. Prigge, G. D. Cole, M. Aspelmeyer, and J. Eisert, Observation of non-Markovian micromechanical Brownian motion, Nat. Commun. {\bf 6}, 7606 (2015).
\bibitem{illuminatiprl} G. Torre, W. Roga, and F. Illuminati, Non-Markovianity of Gaussian Channels, Phys. Rev. Lett. {\bf 115}, 070401 (2015). 
\bibitem{illuminatipra} G. Torre and F. Illuminati, Exact non-Markovian dynamics of Gaussian quantum channels: Finite-time and asymptotic regimes, Phys. Rev. A {\bf 98}, 012124 (2018).
\bibitem{noteOhmic} Note that for strictly Ohmic regime it is assumed the usual bath cut-off $\omega_c$ as the largest energy scale, i.e. $\omega_c \gg \omega_0, \Omega$.

\bibitem{notechi} Due to the perturbative expansion, $\tilde{\gamma}_2(\omega)$ is the only contribution 
from $\tilde{k}_0(\omega)$ 
that enters Eq.~\eqref{kappan},
$\tilde{\gamma}_1(\omega)$ having been discarded.
\bibitem{notejw} We recall that the only case with a purely Markovian dynamics is  a strictly Ohmic spectral function ${\cal J}_1(\omega)=m\gamma_1\omega$ in the classic (high temperature) regime $T_1\gg \omega_0$, where ${\cal L}_1(t)=m\gamma_1 T_1\delta(t)$~\cite{weiss}.
\bibitem{thorwart} M. Thorwart, E. Paladino, and M. Grifoni, Dynamics of the spin-boson model with a structured environment, Chem. Phys. {\bf 296}, 333 (2004).
\bibitem{paladino} E. Paladino, A. G. Maugeri, M. Sassetti, G. Falci, and U. Weiss, Structured environments in solid state systems: crossover from Gaussian to non-Gaussian behavior, Physica E {\bf 40}, 198 (2007).
\bibitem{nazir14} J. Iles-Smith, N. Lambert, and A. Nazir, Environmental dynamics, correlations, and the emergence of noncanonical equilibrium states in open quantum systems, Phys. Rev. A \textbf{90}, 032114 (2014).
\bibitem{aspelmeyer} M. Aspelmeyer, T. J. Kippenberg, and F. Marquardt, Cavity optomechanics, Rev. Mod. Phys. {\bf 86}, 1391 (2014). 
\bibitem{cottet} A. Cottet, M. C. Dartiailh, M. M. Desjardins, T. Cubaynes, L. C. Contamin, M. Delbecq, J. J. Viennot, L. E. Bruhat, B. Doucot, and T. Kontos, Cavity QED with hybrid nanocircuits: from atomic-like physics to condensed matter phenomena, J. Phys.: Condens. Matter {\bf 29}, 433002 (2017).
\bibitem{scigliuzzo} M. Scigliuzzo, A. Bengtsson, J.-C. Besse, A. Wallraff, P. Delsing, and S. Gasparinetti, Primary Thermometry of Propagating Microwaves in the Quantum Regime, Phys. Rev. X {\bf 10}, 041054 (2020).
\bibitem{barzanjeh22} S. Barzanjeh, A. Xuereb, S. Gr\"{o}blacher, M. Paternostro, C. A. Regal, and E. M. Weig, Optomechanics for quantum technologies, Nat. Phys. {\bf 18}, 15 (2022).
\bibitem{sidebandcooling} J. D. Teufel, T. Donner, D. Li, J. W. Harlow, M. S. Allman, K. Cicak, A. J. Sirois, J. D. Whittaker, K. W. Lehnert, and R. W. Simmonds, Sideband cooling of micromechanical motion to the quantum ground state, Nature {\bf 475}, 359 (2011).
\bibitem{note3} A small broadening of the resonances  is due to the damping factors $\gamma_1,\gamma_2$.
\bibitem{Rezek06} Y. Rezek and R. Kosloff, Irreversible Performance of a Quantum Harmonic Heat Engine, New J. Phys. {\bf 8}, 83 (2006).
\bibitem{Kosloff07} R. Kosloff and Y. Rezek, The Quantum Harmonic Otto Cycle. Entropy {\bf 19}, 136 (2017).
\bibitem{Buffoni} L. Buffoni, A. Solfanelli, P. Verruchi, A. Cuccoli, and M. Campisi, Quantum Measurement Cooling, Phys. Rev. Lett. {\bf 122}, 070603 (2019).
\bibitem{notaFW} Indeed, the analysis of more general situations will be the subject of future studies.
\bibitem{notaPC} {All numerical results have been obtained adopting standard linear algebra and series acceleration packages in a in--house developed highly parallel code.} A set of Floquet states with $|\mu|\leq 150$ has been adopted as it proved to be adequate to achieve a relative accuracy of $10^{-3}$ or better in all results.

\bibitem{simone} M. M. M\"uller, R. S. Said, F. Jelezko, T. Calarco, and S. Montangero, One decade of quantum optimal control in the chopped random basis, Rep. Prog. Phys. {\bf 85}, 076001 (2021).
\bibitem{plastina} G. Manzano, F. Plastina, and R. Zambrini, Optimal Work Extraction and Thermodynamics of Quantum Measurements and Correlations, Phys. Rev. Lett. {\bf 121}, 120602 (2018).
\bibitem{noe} P. A. Erdman and F. No\'e, Identifying optimal cycles in quantum thermal machines with reinforcement-learning, NPJ Quantum Inf. {\bf 8}, 1 (2022).
\bibitem{khait2022} I. Khait, J. Carrasquilla, and D. Segal, Optimal control of quantum thermal machines using machine learning, Phys. Rev. Research {\bf 4}, L012029 (2022).
\bibitem{barzanjeh2} S. Barzanjeh, M. Aquilina, and A. Xuereb, Manipulating the flow of thermal noise in quantum devices, Phys. Rev. Lett. {\bf 120}, 060601 (2018).
\bibitem{pontin} A. Pontin, H. Fu, J. H. Iacoponi, P. F. Barker, and T. S. Monteiro, Controlling mode orientations and frequencies in levitated cavity optomechanics, ArXiv:2204.09625 (2022).
\bibitem{viennot} J. J. Viennot, M. C. Dartiailh, A. Cottet, and T. Kontos, Coherent coupling of a single spin to microwave cavity photons, Science {\bf 349}, 408 (2015).
\bibitem{lu21} Y. Lu, A. Bengtsson, J. J. Burnett, E. Wiegand, B. Suri, P. Krantz, A. Fadavi Roudsari, A. F. Kockum, S. Gasparinetti, G. Johansson, P. Delsing, Characterizing decoherence rates of a superconducting qubit by direct microwave scattering, NPJ Quantum Information {\bf 7}, 35 (2021).
\bibitem{rodrigues} D. Bothner, I. C. Rodrigues, and G. A. Steele, Photon-pressure strong coupling between two superconducting circuits, Nat. Phys. {\bf 17}, 85 (2021). 
\bibitem{note_paths} In general, this shift will bring the set $\mathcal{P}(\bar{\mu},n)$ but this does not constitute a problem since the mathematical expressions associated to the sum over all paths are still valid.
\end{thebibliography}
\end{document}